\documentclass[aps,twocolumn,floatfix]{revtex4}
\usepackage{graphicx}
\usepackage{amsfonts}
\usepackage{amsmath,bm,amssymb}
\usepackage{mathtools}
\usepackage{comment}
\usepackage[usenames]{color}
\usepackage{array}
\usepackage[]{changes}

\begin{document}
\DeclarePairedDelimiter\bra{\langle}{\rvert}
\DeclarePairedDelimiter\ket{\lvert}{\rangle}
\DeclarePairedDelimiterX\braket[2]{\langle}{\rangle}{#1 \delimsize\vert #2}

\definechangesauthor[color=red]{JV}
\definechangesauthor[color=blue]{KP}
%\added[id=<id>,remark=<remark>]{text}
%\deleted[id=<id>,remark=<remark>]{text}
%\replaced[id=<id>,remark=<remark>]{newtext}{oldtext}
%\listofchanges[style=<list|summary>]
%remove markup: \usepackage[final]{changes}
%ctan.net has script to remove markup scripts

\title{Magnetic Field Induced Weyl Semimetal from Wannier-Function-based Tight-Binding Model}
\author{John W. Villanova}
\email{jwdvill@vt.edu}
\author{Kyungwha Park}
\email{kyungwha@vt.edu}
\affiliation{Department of Physics, Virginia Tech, Blacksburg, Virginia 24061, USA}

\date{\today}

% projected atomic Wannier functions

\begin{abstract}
% KP: need to revise the first sentence. 04/05/2018
Weyl semimetals (WSMs) have Weyl nodes where conduction and valence bands meet in the absence of inversion or time-reversal symmetry (TRS),
or both. Weyl nodes are topologically protected as long as crystal momentum is conserved, giving rise to Fermi arcs at the surfaces.
Interesting phenomena are expected in WSMs such as the chiral magnetic effect, anomalous Hall conductivity or Nernst effect, and unique
quantum oscillations. The TRS-broken WSM phase can be driven from a topological Dirac semimetal by magnetic field ${\mathbf B}$ or magnetic
dopants, considering that Dirac semimetals have degenerate Weyl nodes stabilized by rotational symmetry, i.e. Dirac nodes, near the Fermi level.
Here we develop a Wannier-function-based tight-binding (WF-TB) model to investigate the formation of Weyl nodes and nodal rings induced by ${\mathbf B}$ field in the topological Dirac semimetal Na$_3$Bi.  The field is applied along the rotational axis. So far, studies of
${\mathbf B}$ field induced WSMs have been limited to cases with $4\times4$ effective models, which may not fully capture interesting effects.
Remarkably, our study based on the WF-TB model shows that upon ${\mathbf B}$ field each Dirac node is split into {\it four} separate Weyl
nodes along the rotational axis near the Fermi level; two nodes with Chern number $\pm 1$ (single Weyl nodes) and two with Chern number $\pm 2$
(double Weyl nodes). This result is in contrast to the common belief that each Dirac node consists of only two Weyl nodes with opposite chirality.
In the context of the $4\times4$ effective models, the existence of double Weyl nodes ensures nonzero cubic terms in momentum.  We further examine the evolution of Fermi
arcs at a side surface as a function of chemical potential.  This analysis corroborates our finding of the double Weyl nodes.  The number of
Fermi arcs at a given chemical potential is consistent with the corresponding Fermi surface Chern numbers.  Furthermore, our study reveals the
existence of nodal rings in the mirror plane below the Fermi level upon ${\mathbf B}$ field.  These nodal rings persist with spin-orbit coupling, in contrast
to many proposed nodal ring/line semimetals.  Our WF-TB model can be used to compute interesting features arising from Berry curvature such as
anomalous Hall and thermal conductivities, and our findings can be applied to other topological Dirac semimetals like Cd$_3$As$_2$.
\end{abstract}
%\pacs{}

\maketitle

\section{Introduction}

% Weyl semimetal in general.

In Weyl semimetals (WSMs), bulk conduction and valence bands touch at an even number of points near the Fermi level, called Weyl nodes, which
are topologically protected by conserved crystal momentum \cite{Nielsen1981,XWan2011}.  The WSM phase occurs when inversion symmetry (IS) or time reversal symmetry (TRS) is broken or both.  A non-zero Chern number is associated with each Weyl node, and it also dictates the number of
open Fermi-arc surface states.  Since early theoretical proposals of WSM in iridates \cite{XWan2011}, HgCr$_2$Se$_4$ \cite{Xu2011,ChenFang2012}, and
TaAs \cite{HWeng2015}, experimental observation of Weyl nodes in IS-broken WSMs TaAs family \cite{BQLv2015,LXYang2015} has stimulated the
field. In addition to Weyl nodes with Dirac dispersion in three orthogonal directions (single Weyl nodes, Chern number $\pm1$), various
types of Weyl nodes were proposed and observed. To name a few, double (triple) Weyl nodes with Chern numbers $\pm2$ ($\pm3$) are associated
with quadratic (cubic) dispersion in the plane orthogonal to the Weyl node separation axis \cite{Xu2011,ChenFang2012,Bernevig2012}; type-II Weyl nodes
\cite{SOLU15,Li2017} are realized when conduction and valence bands meet with the same sign of slope such that electron and hole
pockets are formed near the Fermi level. WSMs are expected to show interesting phenomena arising from Berry curvature, such as the chiral
magnetic effect \cite{SON2012,SON2013,Burkov2015}, anomalous Hall conductivity and Nernst effect \cite{SHARMA16, SHARMA17}, and unique quantum oscillations
\cite{POTT14,MOLL16}.

% TRS-broken WSMs: a smaller number of WSMs along the rotational symmetry axis.

Mostly IS-broken WSMs have been experimentally well characterized, with dozens of Weyl nodes found near the Fermi level, and often off symmetry lines or planes in momentum ${\bf k}$ space \cite{HWeng2015,BQLv2015,LXYang2015}. Experimental studies of TRS-broken WSMs
based on magnetic materials are in debate due to material stability or difficulty in identification of magnetic ordering \cite{HUANG2017}.
%KIYO2016,YANG2017,LIU2017
(Here a strict definition of WSMs is applied where there are no other trivial bands than the bands forming Weyl nodes near the Fermi level.)
A recent study proposed that if doped TRS-broken WSMs with IS can realize a superconducting state with translational symmetry, then
the superconducting state must have an odd-parity, spin-triplet pair potential \cite{Ando2017}. However, such a scenario is not guaranteed
in doped IS-broken WSMs.

One way to induce the TRS-broken WSM phase is to apply a magnetic field ${\mathbf B}$ or insert magnetic dopants in
topological Dirac semimetals (DSMs). Despite TRS and IS, topological DSMs have  degenerate Weyl nodes with opposite chirality, i.e.
Dirac nodes, which are protected by rotational symmetry \cite{BJYang2014}. Topological DSMs Na$_3$Bi and Cd$_3$As$_2$ were experimentally
confirmed to have only two Dirac nodes well separated along the rotational symmetry axis \cite{LIU14_CdAs,LIU14}. Thus, breaking TRS would generate
a much smaller number of Weyl nodes compared to IS-broken WSMs. It is commonly believed that each Dirac node would split into two
Weyl nodes of opposite chirality upon ${\mathbf B}$ field. This arises from studies of the $4\times4$ effective model keeping only
up to quadratic terms in $k$ \cite{Wang2012,BJYang2014,SHARMA16,SHARMA17,WANG13_CdAs}.
Although a possibility of higher-order terms was discussed in the effective model \cite{Wang2012,Gorbar2015,Cano2017}, the existence and strength of such terms have not been investigated before. So far there are no first-principles-based studies of ${\mathbf B}$ field induced WSMs.

In order to investigate topological properties of ${\mathbf B}$ field induced WSMs beyond simple effective models, we develop
a Wannier-function-based tight-binding (WF-TB) model for topological DSM Na$_3$Bi with ${\mathbf B}$ field applied along the rotational
axis. The electronic structure of bulk Na$_3$Bi is first calculated from density-functional theory (DFT) without spin-orbit coupling
(SOC) or ${\mathbf B}$ field. Atom-centered Wannier functions (WFs) are generated from the electronic structure, and we then
construct a WF-TB model by separately adding atomic-like SOC and a Zeeman energy. Landau levels or Peierls phases are not considered
in our WF-TB model. The band structure calculated from the WF-TB model still respects $C_3$ and $6_3$ (screw) symmetries and mirror
symmetry ($\sigma_h$) upon ${\mathbf B}$ field. We avoid maximally-localized WFs in our construction of the WF-TB model.  Topological obstruction~\cite{Thonhauser2006,Soluyanov2011} is not relevant in our WF-TB model since unoccupied bands are included. For example, in topological insulators and semimetals, WF-TB models have been successfully used to investigate topological invariants and other properties \cite{WZhang2010,HWeng2015,Yu2015}.

From the WF-TB model, we find that upon ${\mathbf B}$ field  each Dirac node is split into {\it four} separate Weyl nodes
along the rotational axis near the Fermi level. Two of the nodes have Chern number of $\pm 1$ (single Weyl nodes), while the other
two nodes have Chern number of $\pm 2$ (double Weyl nodes). This result differs from the common belief that each Dirac node consists of
two Weyl nodes of opposite chirality, which is true only when higher-order terms like cubic terms are ignored in the
$4\times4$ effective model. Our calculated Chern numbers associated with the double Weyl nodes unambiguously reveals the existence of
the higher-order terms in momentum.  We further examine the evolution of Fermi arcs at a side
surface as a function of chemical potential, finding that the number of Fermi arcs is consistent with the calculated Chern numbers
associated with the Weyl nodes. This analysis corroborates our findings of the double Weyl nodes.  In addition, our study reveals
that with ${\mathbf B}$ field there are nodal rings in the mirror plane, i.e. $ab$ plane, near the Fermi level. These nodal rings persist
with SOC, in contrast to most proposed nodal ring semimetals where the nodal rings are gapped by SOC except for a few ${\bf k}$ points.
Our WF-TB model with ${\mathbf B}$ field can be used to compute interesting features arising from Berry curvature such as anomalous
Hall and thermal conductivities. Our findings can be applied to other topological Dirac semimetals like Cd$_3$As$_2$.

%The reason lays in the fact that the procedure used in constructing the $4\times4$ effective model from density-functional theory
%(DFT) is not sensitive to the higher-order terms, {\it in the absence of ${\mathbf B}$ field}.

% Outline of the paper for PRB:

 We present crystal structure and symmetries of Na$_3$Bi in Sec.~II and the detailed procedure
of constructing the WF-TB model in Sec.III. Then in Sec. IV we discuss the WF-TB model calculated band structure, the calculated Chern
numbers of the Weyl nodes, the calculated nodal rings, and the evolution of the Fermi arcs versus chemical potential. We conclude in Sec. V.

%From Jack: We confirm the topological charge of the Weyl nodes by calculating the Berry Curvature flux.

\section{Crystal Structure and Symmetries of Na$_3$Bi}
% Space group 194: D_{6h}: 2C6, 2C3, C2, 3C2', 3C2'', i, 2S2, 2S6, sigma_h, 3sigma_d, 3sigma_v.

\begin{figure}[htb]
\centering
\includegraphics[width=0.46 \textwidth]{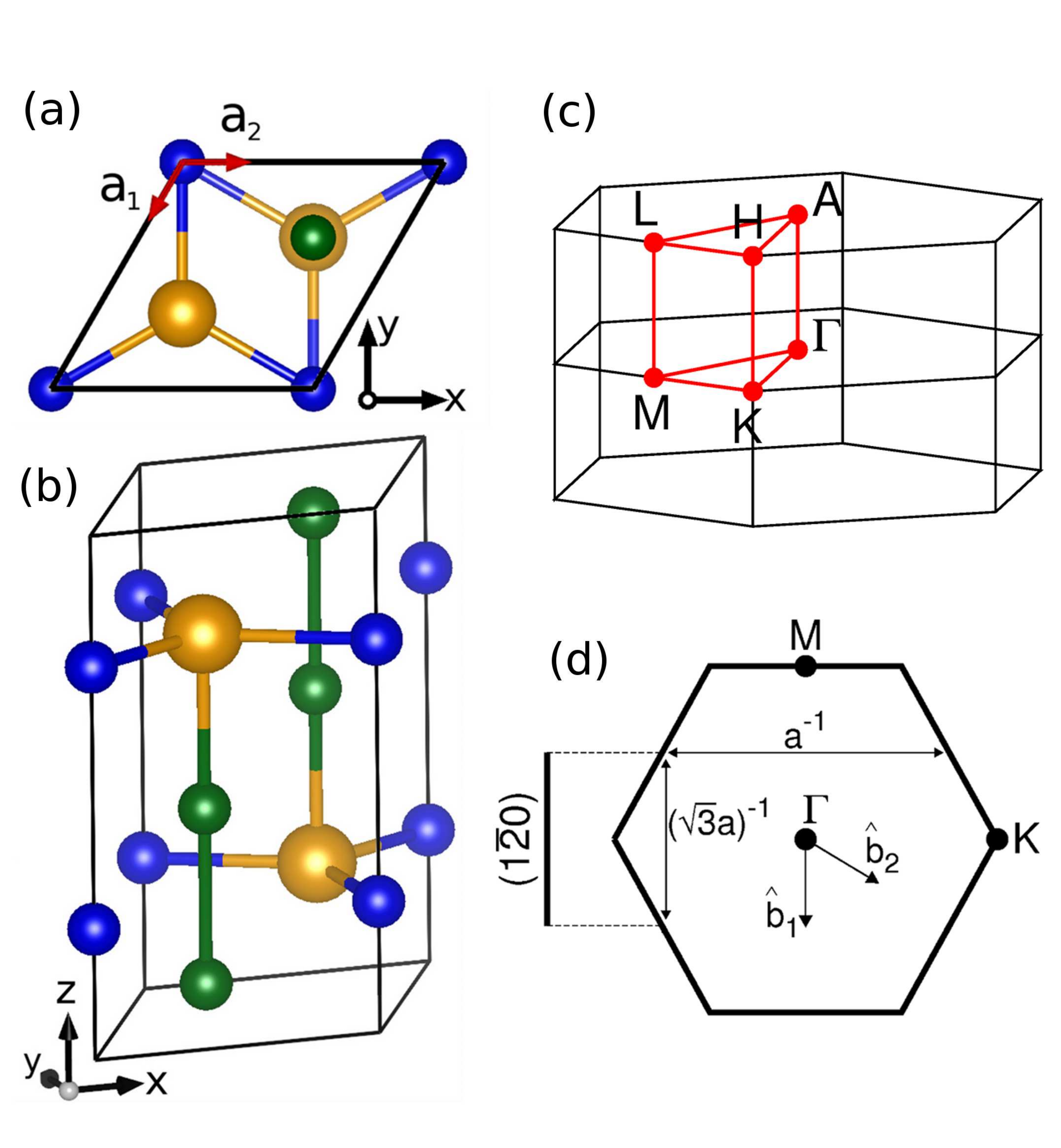}
\caption[geofig]{(Color online) (a) Top-down view of the primitive (001) unit cell of Na$_3$Bi which crystallizes in the P6$_{3}/$mmc hexagonal structure.  Bi atoms are orange and Na atoms are blue (Wyckoff site 2b) and green (4f).  (b) The primitive unit cell at a slight perspective angle.  The in-plane lattice vectors are shown in (a). (c) A first Brillouin zone (BZ)
associated with the primitive (001) unit cell. (d) Top-down view of the primitive cell first BZ showing the projection onto the (1${\bar 2}$0) plane. (Panels (a), (b), and (d) are used with permission, copyright 2017 American Chemical Society \cite{JV2}.)}
\label{fig:geo}
\end{figure}

We consider bulk Na$_3$Bi in space group P6$_3$/mmc (No. 194) with experimental lattice constants of $a=5.448$\ \AA \ and $c=9.655$\ \AA \ \cite{Wang2012} and no geometry relaxation is performed within DFT. There are two inequivalent Na sites, 2b and 4f, with $z=$0.5827 for
the latter in Wyckoff convention, and these are shown in Fig.~\ref{fig:geo} in blue and green, respectively. There is one inequivalent Bi site,
2c in Wyckoff convention shown in gold in Fig.~\ref{fig:geo}. The site symmetries of 2b, 4f, and 2c sites are $D_{3h}$, $C_{3v}$, and
$D_{3h}$, respectively. The primitive unit cell in real space consists of six Na atoms and two Bi atoms, and an associated first
Brillouin zone (BZ) is shown in Fig.~\ref{fig:geo}(c). The bulk crystal has inversion symmetry, $C_3$ rotational and $6_3$ screw symmetries
about the $c$ axis (or $z$ axis), and seven mirror planes (the single horizontal $xy$ plane, $yz$-mirror, and $xz$-glide, as well as the four other planes related to the latter two of these species by rotational symmetry).  Note that our $x$ and $y$ coordinates
are reversed from those in Ref.~\onlinecite{Wang2012}. Henceforth, we refer to the global primitive cell coordinates as unprimed and
the local $(1\bar{2}0)$ cell coordinates as primed. For convenience, we use the $z^{\prime}\perp(1\bar{2}0)$ non-primitive unit cell
for all that follows, unless specified otherwise. The non-primitive unit cell has a rectangular shape and its dimension
is $9.436 \times 9.655 \times 5.448$~\AA$^3$. The volume of the non-primitive unit cell is twice that of the primitive unit cell.
Here the unit vectors in the local coordinates are related to those in the global
coordinates as follows: ($\hat{x}^{\prime},\hat{y}^{\prime},\hat{z}^{\prime}$)$\rightarrow$($-\hat{y},\hat{z},-\hat{x}$).
We consider a side surface $(1\bar{2}0)$ for the study of Fermi-arc surface states in Sec.~IV.D and E.
The reason we use the non-primitive unit cell will be discussed in Sec.~III.B.

\section{Construction of Wannier-function-based tight-binding model}

% We did not include the phase term caused by Peierls instability etc. So not for Landau Levels.

We first calculate the electronic structure of bulk Na$_3$Bi without SOC and ${\mathbf B}$ field by using DFT codes {\sc VASP} \cite{VASP} and
{\sc Quantum Espresso} ({\sc QE}) \cite{QE}. Next we generate the WFs from the DFT-calculated band structure using {\sc Wannier90} \cite{Wannier90}.
Then we construct a tight-binding model from the WFs and add atomic-like SOC and Zeeman energy to the tight-binding model.

\subsection{Initial DFT calculations}

We perform the {\it ab-initio} calculations using {\sc QE} \cite{QE} within the Perdew-Burke-Ernzerhof (PBE)
generalized-gradient approximation (GGA) \cite{PAW, PBE} for the exchange-correlation functional without SOC.  We use the Na.pbe-spn-kjpaw$\_$psl.0.2.UPF and Bi.pbe-dn-kjpaw$\_$psl.0.2.2.UPF projector augmented wave (PAW) pseudopotentials \cite{PSLIB} with an energy cutoff of 50 Ry and smearing of 0.001 Ry.  We consider bulk Na$_3$Bi with the experimental lattice constants \cite{Wang2012} without further relaxation.
We use the $z^{\prime}\perp(1\bar{2}0)$ non-primitive unit cell for all that follows other than the band structure calculation.
We use an $11 \times 11 \times 7$ Monkhorst-Pack $k$-point mesh in the former case and a $7 \times 7 \times 15$ mesh in the latter case.  In both $k$-point samplings, the $\Gamma$ point is included. We also calculate the electronic structure by using {\sc VASP} \cite{VASP} with the
PBE-GGA and PAW pseudopotentials in the absence of SOC and with a $11 \times 11 \times 5$ mesh. We use an energy cutoff of 250~eV and smearing of
0.05~eV. We find excellent agreement between the QE-calculated and the VASP-calculated electronic structures, which justifies our choices of the PAW
pseudopotentials \cite{PSLIB} used in QE.

% (1) definition: (2) what our WFs must satisfy: (3) How to generate them

\subsection{Generation of the Wannier functions}

A Wannier function, $|{\mathbf r}-{\mathbf R}^{\prime}, n\rangle$, centered at position ${\mathbf R}^{\prime}$ in real space is a Fourier
transform of Bloch states, $|\psi_{nk}({\mathbf r})\rangle$, over the ${\mathbf k}$ space, where $n$ is a band index. The Bloch states
can be written as $e^{ i {\mathbf k} \cdot {\mathbf r}} u_{n{\mathbf k}}({\mathbf r})$, where  $u_{n{\mathbf k}}({\mathbf r})$
denotes a lattice-periodic function.
\begin{eqnarray}
|{\mathbf r}-{\mathbf R}^{\prime}, n\rangle &=&
\frac{V}{(2\pi)^3} \int d^3k e^{-i {\mathbf k} \cdot {\mathbf R}^{\prime} } |\psi_{nk}({\mathbf r})\rangle, \\
|\psi_{nk}({\mathbf r})\rangle &=& \sum_{\mathbf R^{\prime}} e^{ i {\mathbf k} \cdot {\mathbf R}^{\prime} }
|{\mathbf r}-{\mathbf R}^{\prime}, n\rangle,
\end{eqnarray}
where $V$ is the volume of the first BZ. Although the concept of Wannier functions was developed very early
\cite{WannierOG}, their practical usage has rapidly developed in the past twenty years since two bottlenecks were removed by Marzari, Vanderbilt,
and collaborators \cite{Marzari1997,Souza2001}. The first difficulty was non-uniqueness of WFs due to gauge freedom, which was resolved
by searching for maximally localized Wannier functions \cite{Marzari1997}. The second bottleneck was dealing with cases in which the set of bands of interest is not separated from a larger set of bands by a gap at every $k$ point, as is the case in metals. This was solved by minimizing the gauge invariant part of the spread functional
\cite{Souza2001}. These features are implemented in {\sc Wannier90} code \cite{Wannier90}.

%%%%% Among the following criteria, categorize essential ingredients and non-essential ingredients??

\begin{figure}[htb]
\centering
\includegraphics[width=0.46 \textwidth]{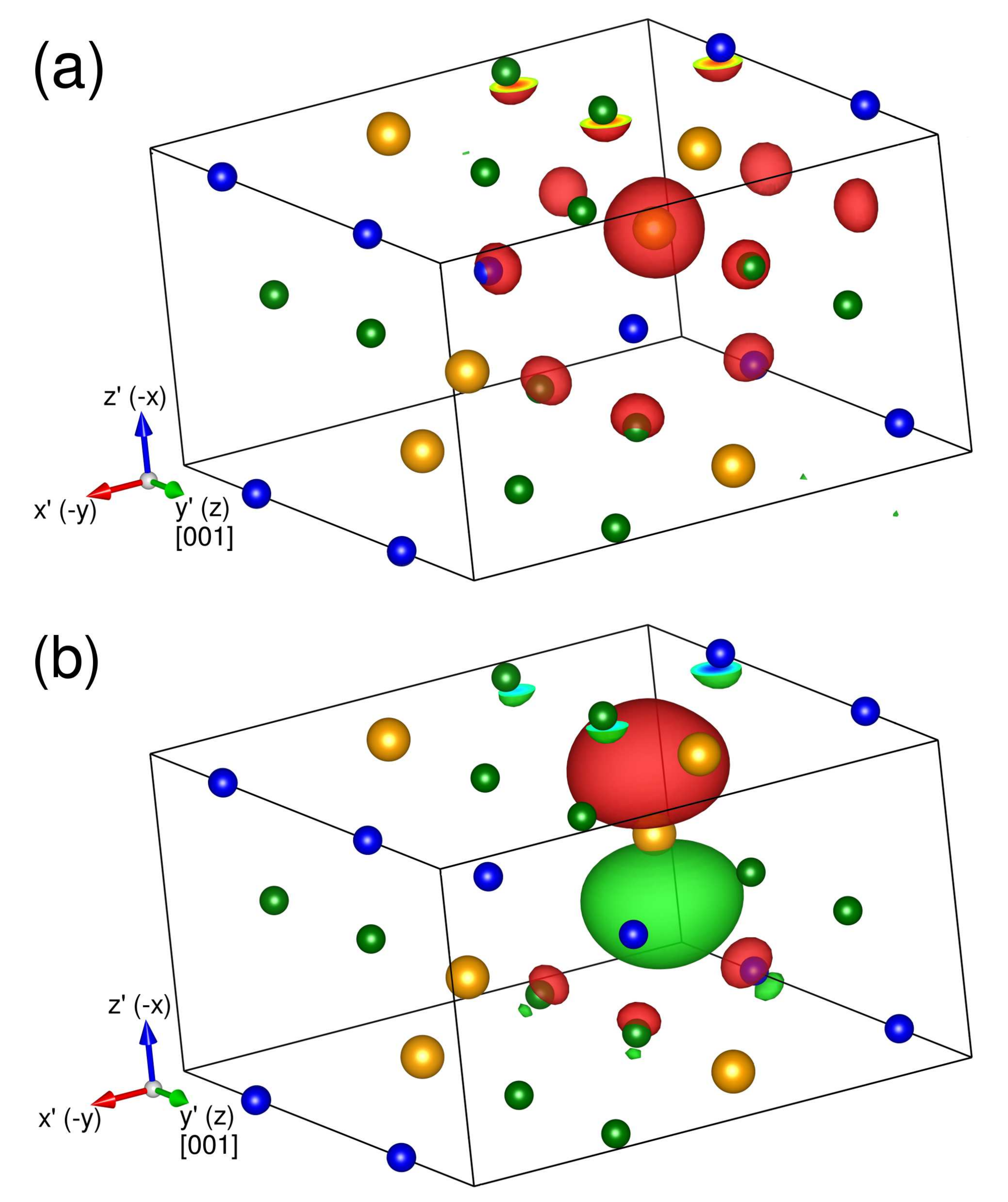}
\caption[$s$- and $p$-orbital Wannier functions in the (1$\bar{2}$0) cell]{(Color online) (a) Wannier function corresponding to an $s$-orbital, and (b) Wannier function corresponding to a $p_{z^{\prime}}$-orbital in the non-primitive unit cell.  An isosurface level of 3 is chosen in both panels.  Bi atoms are orange and Na atoms are blue (Wyckoff site 2b) and green (4f).  Green Na atoms are vertically oriented relative to the Bi atoms in the primitive hexagonal cell.}
\label{fig:WFs}
\end{figure}

Before presenting our WF generation, let us discuss several criteria that the WFs must satisfy in our WF-TB model. First, the WF-TB
model must reproduce the DFT-calculated band structure near the Fermi level with and without SOC. Some occupied and unoccupied bands near the
Fermi levels are needed to investigate the Dirac and Weyl nodes. Second, the band structure obtained from the WF-TB model must respect inherent
crystal symmetries with and without ${\mathbf B}$ field. There is a topological obstruction \cite{Thonhauser2006,Thouless1984,Soluyanov2011} to the construction of (maximally-localized) WFs for Chern insulators and topological insulators when only occupied bands are considered. In addition, maximally localized WFs tend to break crystal symmetries, while projected atomic WFs (without maximal localization) respect crystal symmetries ~\cite{WZhang2010}. Thus, we carry out only the disentanglement procedure by minimizing the gauge-invariant part of the spread
functional $\Omega_I$. The definition of $\Omega_I$ can be found from Ref.~\onlinecite{Souza2001}. Third, the Wannier centers should be at
the atomic centers. Fourth, the WFs must be either very close to pure atomic orbitals or a linear combination of them. The third and fourth
criteria are required because we add the atomic-like SOC and a Zeeman energy term. Fifth, the spread functional of each WF must not be too
large to exceed the size of the home cell. Sixth, a small number of WFs is desirable in order to reduce the size of the Hamiltonian matrix.

In order to first construct the SOC-free WF-TB model, we start with an initial set of 16 projected atomic orbitals, $g_{j}$, comprising
$s$-, $p_{x^{\prime}}$-, $p_{y^{\prime}}$-, and $p_{z^{\prime}}$-orbitals centered at the four Bi atoms in the
$z^{\prime}\perp(1{\bar 2}0)$ unit cell.  Using the DFT-calculated Bloch eigenvalues and eigenstates, we compute the overlap matrix,
$M_{mn}^{{\mathbf k},{\mathbf b}}=\langle u_{m{\mathbf k}}|u_{n{\mathbf k}+{\mathbf b}}\rangle$, and projection matrix,
$A_{mj}^{({\mathbf k})}=\langle\psi_{m{\mathbf k}}|g_j\rangle$, at each DFT-sampled $k$ point by using {\sc Wannier90} \cite{Wannier90}.
Here $m$ and $n$ are indices of the Bloch states or bands, ${\mathbf b}$ is a vector between two neighboring $k$ points, and $j$ is the
WF index. Then we apply only the disentanglement procedure within the outer energy window $[-3.86,5.44]$~eV with respect to the
Fermi level. In this energy window, the number of Bloch bands ($27 \le N_b({\bf k}) \le 34$) is much greater than the number of the WFs,
where the number of occupied bands is 12 and the total number of Bloch bands depends on ${\bf k}$. We find that the generated WFs have
only real components and that the WFs are close to pure atomic orbitals, as shown in Fig.~\ref{fig:WFs}. In the case of the projected
$s$-orbital WFs, there are small contributions from the neighboring Na sites. This does not affect our implementation of SOC
since the $s$-orbital WFs do not contribute to SOC. The Wannier centers are at the atomic centers within the order of
0.001~\AA~for the $p$-orbital WFs and 0.01~\AA~for the $s$-orbital WFs. The spreads of the individual WFs as well as the Wannier centers
are listed in Table~\ref{tab:120wfs}. The $p$-orbital WFs are  well localized, whereas the $s$-orbital WFs are substantially delocalized
but their spreads remain within the home non-primitive unit cell. Such spreads could be the reason we obtain a better set of WFs when we
increase the unit cell size in real space, compared to the case of using a primitive unit cell. Here a ``better set" of WFs means improved
agreement with the DFT-calculated band structure while respecting the crystal symmetries.
A similar effect has been discussed for topological insulators \cite{Mustafa2016}. With the generated WFs, the gauge-invariant
part of the spread functional $\Omega_I$ is 84.69~\AA$^2$, and the diagonal and off-diagonal non-invariant part of the spread functionals
$\Omega_D$ and $\Omega_{OD}$ are 0.15 and 19.21~\AA$^2$, respectively. We also check that matrix
$A^{\dag ({\mathbf k})}A^{({\mathbf k})}$ is not singular at any ${\bf k}$ points for our choice of the initial set $g_j$.

\setlength{\tabcolsep}{14pt}
\begin{table*}[htb]
\centering
\caption[WFs for the (1${\bar 2}$0) cell]{Cartesian positions (\AA) of the WFs for the (1${\bar 2}$0) cell with their spreads.
The atomic positions are listed at the top of the table for ease of comparison.  All coordinates are \emph{local}.}
\begin{tabular}{c c c c c c}
\hline\hline
Atom & Orbital & $x^{\prime}$ & $y^{\prime}$ & $z^{\prime}$ & Spread (\AA$^2$) \\
\hline
Bi 1 & & 1.57270 &  2.41375 &  2.72400 \\
Bi 2 & & 3.14540 &  7.24125 &  0.00000 \\
Bi 3 & & 6.29081 &  2.41375 &  0.00000 \\
Bi 4 & & 7.86351 &  7.24125 &  2.72400 \\ [0.5ex]
\hline
1 & $s$  & 1.53021 & 2.41377 & 2.72400 & 13.30 \\
2 & $s$  & 3.18791 & 7.24123 & 0.00000 & 13.30 \\
1 & $p_{z^{\prime}}$ & 1.56990 & 2.41375 & 2.72400 & 4.36 \\
1 & $p_{x^{\prime}}$ & 1.57225 & 2.41375 & 2.72400 & 4.36 \\
1 & $p_{y^{\prime}}$ & 1.56931 & 2.41375 & 2.72400 & 3.99 \\
2 & $p_{z^{\prime}}$ & 3.14821 & 7.24125 & 0.00000 & 4.36 \\
2 & $p_{x^{\prime}}$ & 3.14587 & 7.24125 & 0.00000 & 4.36 \\
2 & $p_{y^{\prime}}$ & 3.14880 & 7.24125 & 0.00000 & 3.99 \\
3 & $s$  & 6.24840 & 2.41377 & 0.00000 & 13.30 \\
4 & $s$  & 7.90603 & 7.24123 & 2.72400 & 13.30 \\
3 & $p_{z^{\prime}}$ & 6.28801 & 2.41375 & 0.00000 & 4.36 \\
3 & $p_{x^{\prime}}$ & 6.29034 & 2.41375 & 0.00000 & 4.36 \\
3 & $p_{y^{\prime}}$ & 6.28742 & 2.41375 & 0.00000 & 3.99 \\
4 & $p_{z^{\prime}}$ & 7.86632 & 7.24125 & 2.72400 & 4.36 \\
4 & $p_{x^{\prime}}$ & 7.86398 & 7.24125 & 2.72400 & 4.36 \\
4 & $p_{y^{\prime}}$ & 7.86691 & 7.24125 & 2.72400 & 3.99 \\
\hline
\end{tabular}
\label{tab:120wfs}
\end{table*}

\subsection{Spin-free Hamiltonian}

Now we construct the spin-free WF-TB model, by using the generated WFs, $|{\mathbf R}+{\mathbf s}_{\beta}\rangle$, centered at
${\mathbf R}+{\mathbf s}_{\beta}$, where ${\mathbf R}$ are the lattice vectors and ${\mathbf s}_{\beta}$ denote the sites of
orbital ${\beta}$ ($\beta$$=$1,...,16). The spin-free Hamiltonian matrix ${\cal H}_0$ reads
\begin{eqnarray}
{\cal H}_{0,\alpha \beta}({\mathbf k}) &=& \langle \psi_{{\mathbf k},\alpha}| {\cal H}_0 |\psi_{{\mathbf k},\beta}\rangle,  \\
 &=& \sum_{{\mathbf R}} e^{-i {\mathbf k} \cdot ({\mathbf R}+{\mathbf s}_{\alpha}-{\mathbf s}_{\beta})}
 t_{\alpha \beta}({\mathbf R}-{\mathbf 0}), \label{eq:Hab} \\
t_{\alpha \beta}({\mathbf R}-{\mathbf 0}) &=& \langle {\mathbf R} + {\mathbf s}_{\alpha}| {\cal H}_0 | {\mathbf 0} + {\mathbf s}_{\beta} \rangle,
\end{eqnarray}
where $t_{\alpha \beta}({\mathbf R}-{\mathbf 0})$ is a hopping or tunneling parameter from orbital $\beta$ at site ${\mathbf s}_{\beta}$
in the home cell at ${\mathbf R}={\mathbf 0}$ to orbital $\alpha$ at site ${\mathbf s}_{\alpha}$ in the unit cell located at ${\mathbf R}$.
Note that we have four different Bi sites within the non-primitive unit cell. The factor $e^{-i {\mathbf k} \cdot ({\mathbf s}_{\alpha}-{\mathbf s}_{\beta})}$ in Eq.~(\ref{eq:Hab}) can be absorbed into a new basis set.

\subsection{Addition of spin-orbit coupling and Zeeman term}

Since the atom-centered Wannier functions are very close to pure states of the orbitals we project onto, on-site SOC is added to the home-cell
terms directly.  The matrix form of SOC in the basis set of $\{\ket{s,\uparrow},\ket{p_{z^{\prime}},\uparrow},\ket{p_{x^{\prime}},\uparrow},\ket{p_{y^{\prime}},\uparrow},
\ket{s,\downarrow},\ket{p_{z^{\prime}},\downarrow},\ket{p_{x^{\prime}},\downarrow},\ket{p_{y^{\prime}},\downarrow}\}$ for a single Bi atom is
\begin{equation}
{\cal H}_{\small SOC} = \lambda \mathbf{L}\cdot {\mathbf{\sigma}} =
 \frac{\lambda}{2}\left(\begin{array}{cccccccc}
 0 & 0 & 0 & 0 & 0 & 0 & 0 & 0 \\
 0 & 0 & 0 & 0 & 0 & 0 & -1 & i \\
 0 & 0 & 0 & -i & 0 & 1 & 0 & 0 \\
 0 & 0 & i & 0 & 0 & -i & 0 & 0 \\
 0 & 0 & 0 & 0 & 0 & 0 & 0 & 0 \\
 0 & 0 & 1 & i & 0 & 0 & 0 & 0 \\
 0 & -1 & 0 & 0 & 0 & 0 & 0 & i \\
 0 & -i & 0 & 0 & 0 & 0 & -i & 0 \\
 \end{array} \right),
\end{equation}
where $\lambda$ is the SOC parameter, ${\mathbf L}$ is the orbital angular momentum, and ${\mathbf {\sigma}}$ represent Pauli spin matrices.  With our generated WFs, we now have a $32 \times 32$ matrix ${\cal H}_{\small SOC}$ since there are four Bi sites per non-primitive unit cell. We find that $\lambda=1.165$ eV reproduces the DFT band structure around the Fermi level the best,
which is to be favorably compared to $\lambda_{\rm Bi}=1.25$ eV \cite{ZHANG_NJP}.

We add the magnetic field as a Zeeman term ${\cal H}_{Z} = \tilde{g}\mu_{B}(\mathbf{L}+2\mathbf{S})\cdot \mathbf{B}$, where $\mu_B$ is Bohr 
magneton and ${\mathbf S}$ is the spin angular momentum. For a free electron, $\tilde{g}=1$. 
We do not include Peierls phases in the hopping parameters $t_{\alpha \beta}$.
The total Hamiltonian for the WF-TB model is ${\cal H} = {\cal H}_0 + {\cal H}_{\small SOC} + {\cal H}_Z$.
For the results presented through the rest of this work, we consider $\tilde{g}\mu_{B}B_{z}$$=$ 0.025~eV, unless specified otherwise.
The experimentally realized Dirac semimetals exhibit large g-factors, with $g$~$\approx$~20 in Na$_3$Bi \cite{XION15} and $g$~$\approx$~40 in Cd$_3$As$_2$ \cite{JEON14}, and the magnetic field strength that we consider is experimentally achievable. However, our findings would {\it not} qualitatively change with the field strength if
the magnetic field is not extremely high. For example, the number of the Weyl nodes near the Fermi level, the Chern numbers associated with the
Weyl nodes, the existence of the cubic terms in $k$ in the $4 \times 4$ effective model, and the existence of at least one nodal ring,
do not change when $\tilde{g}\mu_{B}B_{z}$ is less than 0.05~eV. We choose the particular value of ${\mathbf B}$ field since the splitting of each Dirac node is more visible and easier to analyze.

\section{Results and Discussion}

% 1) TB model shows the correct band structure and the correct symmetry expected. Justification that our TB model is good enough to study Weyl nodes.
% 2) Identification of Weyl nodes by computing topological charges of the Berry curvature flux associated with the band crossing points.
%    Estimate of the higher-order terms.
% 3) Identification of Fermi arcs and rationalize their numbers based on the calculated Chern numbers.
% 4) Berry curvature (global z component) magnitude as a function of kz and ky at kx=0 plane + Anomalous transverse Hall and thermal conductivities.

\subsection{Calculated band structure and symmetry}

\begin{figure*}[htb]
\begin{center}
\includegraphics[width=0.85 \textwidth]{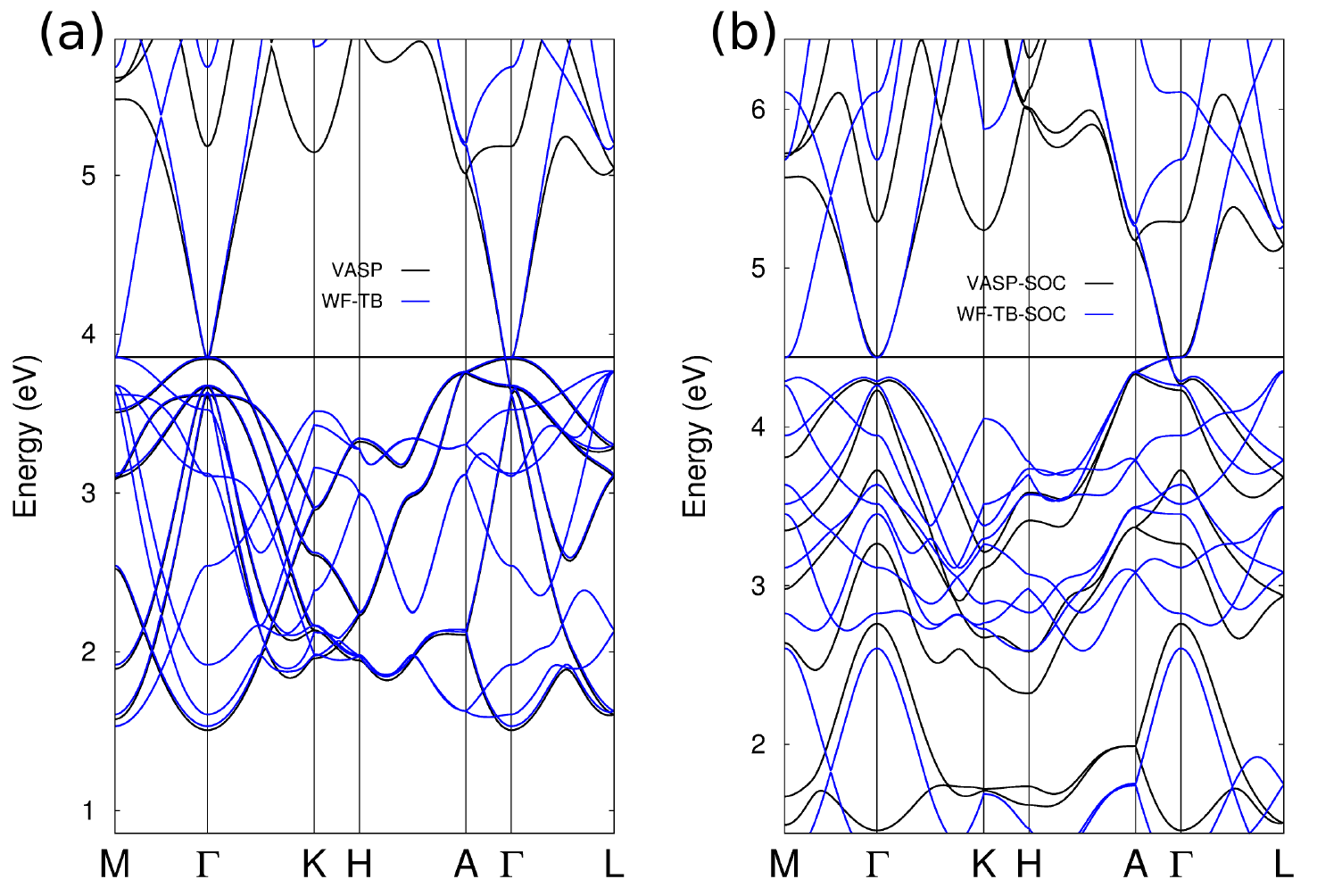}
\caption[a figure]{(Color online) (a) Comparison of the bulk band structure without SOC of Na$_3$Bi, using VASP with the primitive unit cell versus the
WF-TB Hamiltonian with the non-primitive unit cell. (b) Likewise with SOC.  In both cases the band-folding is apparent, especially where the
(1${\bar 2}$0)-BZ ends halfway along $\text{M}\Gamma$. The horizontal solid lines in the middle in (a) and (b) indicate the Fermi levels.
Here (a) and (b) are without ${\mathbf B}$ field.}
\label{fig:bulk_120}
\end{center}
\end{figure*}

\begin{figure*}[htb]
\begin{center}
\includegraphics[width=0.9 \textwidth]{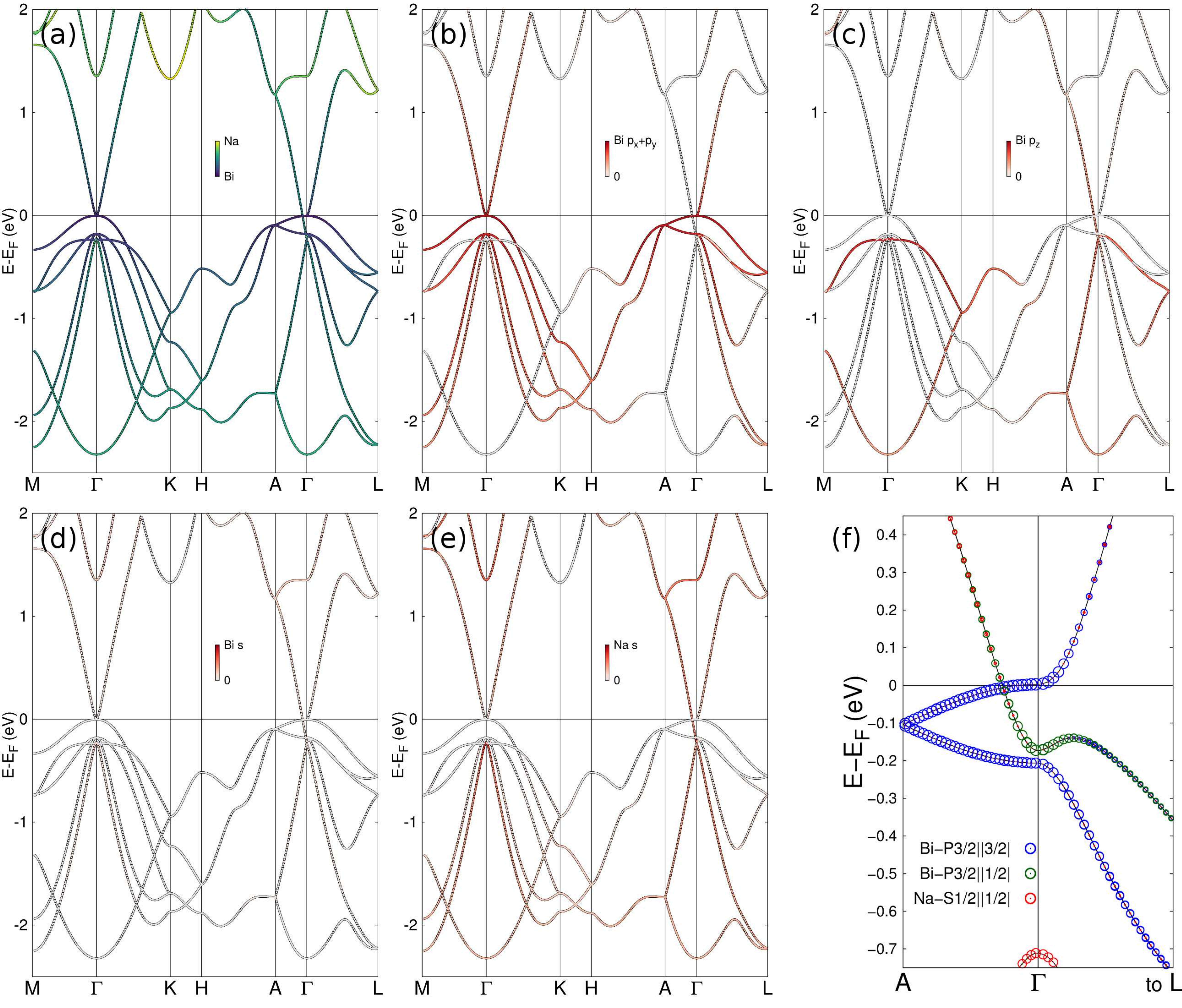}
\caption[a figure]{(Color online) (a) QE-calculated band structure without SOC colored for the contributions of Na and Bi states.  (b-e) QE-calculated band structure without SOC colored for the fraction of the contribution of each orbital at each $k$ point, (b) Bi in-plane $p_x$ and $p_y$ orbitals, (c) Bi $p_z$, (d) Bi $s$, and (e) Na $s$. (f) QE-calculated band structure \emph{with} SOC, demonstrating the Dirac node as a crossing of two doubly-degenerate bands with different rotational eigenvalues, $|j_z|=3/2$ (blue) and $|j_z|=1/2$ (red and green).  All cases use the primitive unit cell without ${\mathbf B}$ field.}
\label{fig:bulkproj}
\end{center}
\end{figure*}

We check that our WF-TB model reproduces the first-principles band structure without SOC.
Figure~\ref{fig:bulk_120}(a) shows the WF-TB-calculated band structure overlain with the VASP-calculated one in the absence of SOC. Except
for the band-folding, there is excellent agreement between the two band structures over a wide range of energies, approximately within
[-4.0, 1.0]~eV with respect to the Fermi level. In the Appendix, we show both the VASP-calculated and the WF-TB-calculated band structures in
the non-primitive unit cell for a comparison with band-folding.
Especially, the band structure near the Dirac node along the A-$\Gamma$ direction is well reproduced from the WF-TB model Hamiltonian.  We note that the VASP-calculated band structure is identical to the QE-calculated one. In the vicinity of the Dirac node, the QE-calculated band structure demonstrates that the two crossing bands, without SOC, consist of one band with Na $s$, Bi $s$, and Bi $p_z$ orbital characters, and the other band with Bi $p_x$ and $p_y$ orbital characters (in the global coordinates).
The composition of the orbital characteristics is shown in Fig.~\ref{fig:bulkproj}. The contribution of the Na $s$ orbital is hardly larger than the contribution of the Bi $s$ plus $p_z$ orbitals along the A-$\Gamma$ direction. This fact, along with the observation that the resultant WFs comport with our six criteria, justifies our choice of the initial set, $g_j$.

\begin{figure}[htb]
\begin{center}
\includegraphics[width=0.48 \textwidth]{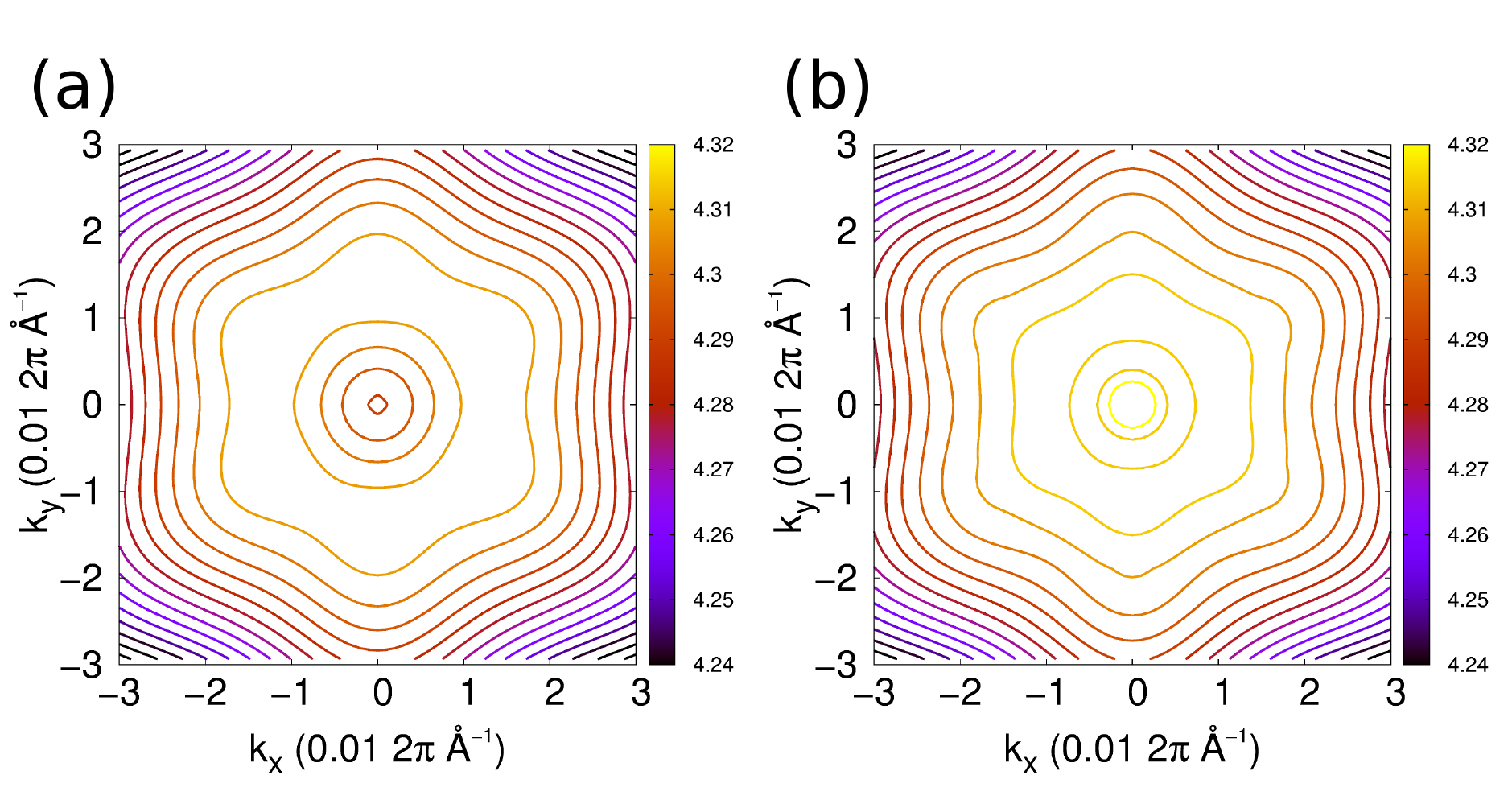}
\caption[a figure]{(Color online) (a) Constant energy contours of the top valence band in the $k_z$$=$0 plane as calculated from the WF-TB model for the $(1\bar{2}0)$ cell in the absence of the magnetic field. (b) Constant energy contours of the top valence band in the $k_z$=0 plane as calculated from the WF-TB model of the (1$\bar{2}$0) cell with an applied magnetic field of 25 meV, exhibiting the hexagonal symmetry.  In both panels, SOC is included and the colorscale is for the energy (eV). The Fermi level in (a) and (b) is 4.44~eV.}
\label{fig:cec120}
\end{center}
\end{figure}

Now we diagonalize the $32 \times 32$ matrix, ${\cal H}_0+{\cal H}_{\small SOC}$, with the same interpolated $k$ points, in order to check if
our WF-TB model reproduces the first-principles band structure with SOC. As shown in Fig.~\ref{fig:bulk_120}(b), with SOC, we also find
excellent agreement between the WF-TB- and VASP-calculated band structures near the Fermi level. See also the band structures in
the non-primitive unit cell in the Appendix for a comparison with band-folding.
The Dirac node from the WF-TB model is found to be at 0.08785~\AA$^{-1}$, which agrees well with the location of the DFT-calculated Dirac node.  Furthermore, we investigate the symmetry of the WF-TB-calculated
band structure by computing the constant energy contours of the top valence band in the global-$xy$ plane.  With and without magnetic field, we find sixfold rotational symmetry (Fig.~\ref{fig:cec120}). This result is consistent with the $6_3$ screw and $\sigma_h$ (mirror symmetry about the
$xy$ or $z^{\prime}x^{\prime}$ plane) crystal symmetries, adding further credence to the validity of the WF-TB model. Note that we carefully
check all aspects of the forthcoming results near the Fermi level, confirming that those results are not influenced by the band folding.

\subsection{Splitting Dirac Nodes via Magnetic Field: Single and Double Weyl nodes}

% our determination of chirality

\begin{figure*}[htb]
\begin{center}
\includegraphics[width=0.7 \textwidth]{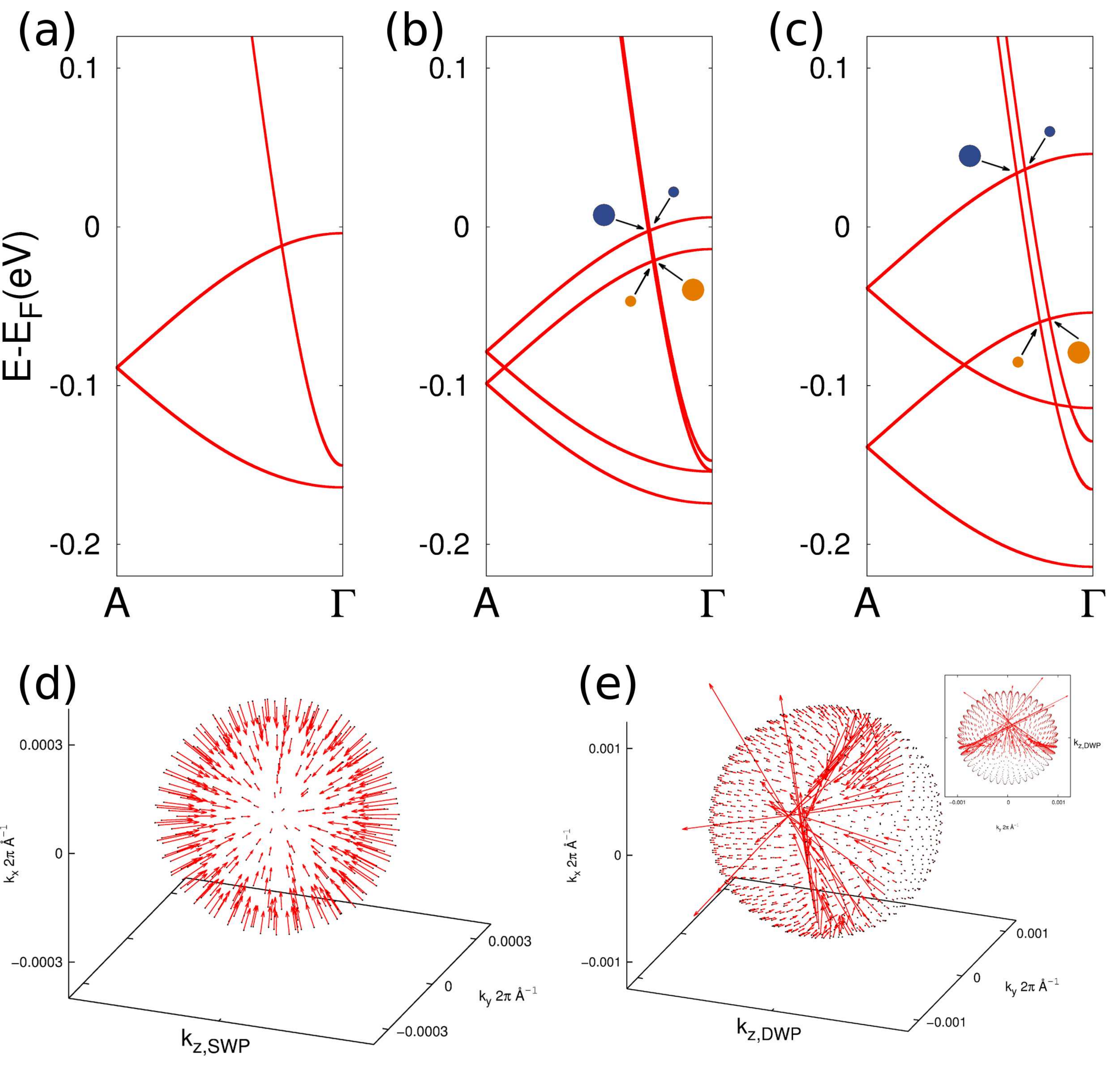}
\caption[Band structure in an applied magnetic field]{(Color online) Band structure of Na$_3$Bi along the rotation axis (a) without ${\mathbf B}$ field,
(b) with $\tilde{g}\mu_B B_z$$=$5 meV, and (c) with $\tilde{g}\mu_B B_z$$=$25 meV.   In (b) and (c)
the positive (negative) chirality Weyl nodes are noted in orange (blue) circles, with the small (large) circles meaning Chern numbers
of $\pm 1$ ($\pm 2$). (d) Berry curvature in the vicinity of a single Weyl point $W_{25}$ with chiral charge of $-1$. The ``anti-hedgehog" monopole
shape is apparent. (e) Berry curvature in the vicinity of a double Weyl point $W_{24,a}$ with chiral charge of $-2$. The ``anti-hedgehog"
shape is quite different, with almost all of the Berry curvature arising on a chord of the sphere.  The inset shows the projection of the Berry
curvature of (e) onto the $k_z=0$ plane, showing the chord which is almost along an equator of the sphere.}
\label{fig:nomuss}
\end{center}
\end{figure*}

We apply a magnetic field along the $6_3$ screw axis ($c$ axis) of Na$_3$Bi, finding that each Dirac node [Fig.~\ref{fig:nomuss}(a)] splits into four separate Weyl nodes along the $c$ axis.  Figure \ref{fig:nomuss}(b) and (c) shows the development of the four Weyl nodes along
the $c$ axis in the half-BZ upon ${\mathbf B}$ field.  Between these values of ${\mathbf B}$ field, the number of field-induced Weyl nodes does not change and nor do the chiralities of the nodes.  We henceforth present only the result for $\tilde{g}\mu_B B_z$$=$0.025 eV, though we note that there is still no qualitative change in the results even for larger fields $\tilde{g}\mu_B B_z$$<$0.05 eV.
Each Weyl node is labeled such that $W_{n}$ denotes a band crossing point between bands $n$ and $n+1$.  When there are multiple crossing
points arising from bands $n$ and $n+1$, an additional index is included right next to the band index. In order to determine the chiral charge of the Weyl nodes, we first calculate the Berry curvature of the Bloch bands obtained from our generated WF-TB model including the SOC and Zeeman term,
by using {\sc Wannier Tools} \cite{WTOOLS}.  The Berry curvature of band $n$ in momentum space, ${\mathbf \Omega}_n({\mathbf k})$, is defined to be
${\mathbf \nabla}_{\mathbf k} \times {\mathbf A}_n({\mathbf k})$, where
${\mathbf A}_n({\mathbf k})=i \langle u_{n{\mathbf k}}|{\mathbf \nabla_{\mathbf k}}u_{n{\mathbf k}}\rangle$.
Defining ${\cal H}_{\alpha}\equiv \partial {\cal H} \slash \partial k_{\alpha}$ and within the space represented by the WFs, the Berry curvature can be calculated as \cite{XWang2006}
\begin{equation}
\epsilon_{\alpha\beta\gamma}\Omega_{n,\gamma}= -2 {\rm Im} \sum_{m \neq n} \frac{\langle\langle \phi_n \|{\cal H}_{\alpha}\|\phi_m \rangle\rangle
\langle\langle \phi_m \|{\cal H}_{\beta}\|\phi_n \rangle\rangle}{({\cal E}_m - {\cal E}_n)^2},
\end{equation}
where $\|\phi_n \rangle\rangle$ and ${\cal E}_n$ are the $n$-th eigenvector and eigenvalue of ${\cal H}$ (the $32\times32$ Hamiltonian matrix
discussed in Sec. III), and $\epsilon_{\alpha\beta\gamma}$ is the Levi-Civita symbol, though no sum over $\gamma$ is implied.
%Further, ${\cal \bar{H}}^{(\text{H})}_{\alpha}=U^{\dagger}{\cal H}_{\alpha}U$ and $U$ is the matrix of column-eigenvectors which diagonalizes ${\cal H}$.
Then we calculate the Chern number or Berry curvature flux $\chi_n$ of each Weyl node $W_n$ by enclosing it in spheres of successively smaller radius,
\begin{equation}
\chi_n = \frac{1}{2\pi}\oint_{S} dS \ \hat{\mathbf{n}} \cdot \mathbf{\Omega}_n(\mathbf{k}),
\label{eqn:chirality}
\end{equation}
where $S$ is the two-dimensional surface of the sphere, or, as relevant later, $S_n$ is the Fermi surface (FS) sheet of band $n$,
and ${\mathbf{n}}$ is a unit vector normal to $S$ or $S_n$.
Our calculation shows that the four Weyl nodes consist of two Weyl nodes with $\chi_n=\pm1$ and two nodes with $\chi_n=\pm2$. The former
(latter) nodes are referred to as single (double) Weyl nodes.  The Berry curvature vector fields for the single and double Weyl nodes are shown
in Fig.~\ref{fig:nomuss}(d) and (e).  The calculated Chern numbers agree with the expected dispersion around the nodes.  The bands forming
the single Weyl nodes disperse linearly in all directions around the nodes. For the double Weyl nodes, the bands disperse linearly along the
$z$-axis (rotational axis) and quadratically in the $xy$-plane.  The positions, energies, and chiralities of the Weyl nodes are listed in Table \ref{tab:WNs}. One pair of single and double Weyl nodes, $W_{25}$ ($\chi_{25}=-1$) and $W_{24,a}$ ($\chi_{24,a}=-2$),  are located at higher energies, while the other two nodes, $W_{23}$ ($\chi_{23}=+1$) and $W_{24,b}$ ($\chi_{24,b}=+2$),  are found at lower energies.

\setlength{\tabcolsep}{14pt}
\begin{table*}[htb]
\centering
\caption[Momenta and energies of the Weyl nodes]{Chern number, momentum, and energy with respect to the Fermi level of the four Weyl nodes in the $z>0$ half-BZ, where $W_{n,\alpha}$ is $\alpha$-th Weyl node arising from band $n$, the corresponding band index shown in Fig.~\ref{fig:FASSBZ}(a) and used in Eq.~\ref{eqn:chernFS}. Note that $\tilde{g}\mu_B B_z$$=$0.025 eV.}
\begin{tabular}{c c c c c c c}
\hline\hline
$W_{n,\alpha}$ & $\chi_n$ & $k_z$ (\AA$^{-1}$) & Energy (meV) \\
\hline
$W_{24,a}$ & $-$2 & 0.10973 & 33.3 \\
$W_{25}$ & $-$1 & 0.09779 & 35.8 \\
$W_{23}$ & +1 & 0.07618 & -60.4 \\
$W_{24,b}$ & +2 & 0.06237 & -58.5 \\
\hline
\end{tabular}
\label{tab:WNs}
\end{table*}

% comparison + analysis below: Hongming's case considered double Weyl nodes. Gorbar did not. Our estimate of cubic terms.
% Comparison with Kargarian and the recent beta-CuI paper.

\begin{figure}[htb]
\begin{center}
\includegraphics[width=0.42\textwidth]{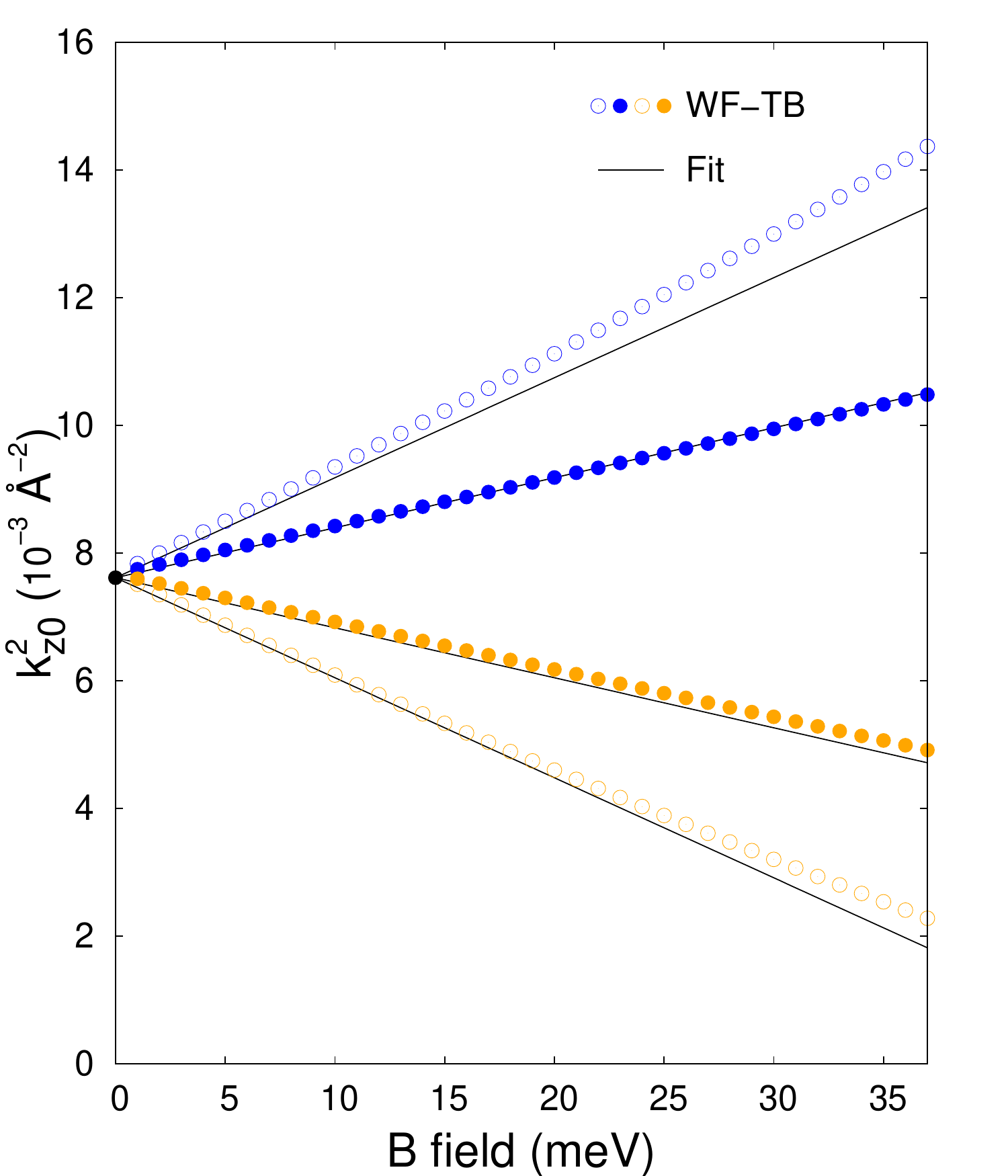}
\caption[kz0 fitting]{(Color online) Evolution of the position of the Weyl nodes calculated with the WF-TB model versus magnetic field strength.  The single (double) Weyl node positions are shown as closed (open) circles, and the blue (orange) color corresponds to the negative (positive) chiral charge.  The solid black lines correspond to the average of four fits which demonstrate good agreement with the $4 \times 4$ Hamiltonian, especially at low fields. The fit for the single Weyl node ($\chi_{25}$=$-$1) is $k_{z0}^2(B_z) = (7.613 \cdot 10^{-3} $\AA$^{-2})+(7.830\cdot 10^{-3}~($eV$\cdot$\AA$^2)^{-1})\cdot B_z[$eV$]$.  The other fitting curves are given by the appropriate choice of the sign of the slope, and the double Weyl node fits have twice the magnitude of the slope.}
\label{fig:fitting}
\end{center}
\end{figure}

Each Weyl node is the result of a crossing of bands with different rotational eigenvalues, inherited from the Dirac node in the absence of the
magnetic field.  For example, band crossings between $j_z=\pm3/2$ and $j_z=\mp1/2$ bands create double Weyl nodes, while band crossings between $j_z=\pm3/2$ and $j_z=\pm1/2$ produce single Weyl nodes, where $j_z$ is the eigenvalue of the $z$ component of the total angular momentum
operator ${\mathbf J}$, as discussed in Refs.~\onlinecite{Wang2012,Cano2017}.  We confirm that the $j_z$ values for the crossing bands obtained
from the WF-TB model agree with this analysis.  The 6$_3$ screw symmetry allows double Weyl nodes \cite{Vanderbilt_Te}.  
Figure~\ref{fig:fitting} shows the field-dependence of the Weyl-node positions $k_{z0}$ along the $k_z$ axis calculated with the WF-TB model.
At low ${\mathbf B}$ fields the $k_{z0}^2$ values evolve linearly with field, which is consistent with the Weyl-node positions obtained
from the effective $4 \times 4$ model (shown below) \cite{Wang2012,Cano2017}. Although the single Weyl nodes induced by ${\mathbf B}$ field 
were reported in the literature \cite{Wang2012,Gorbar2015}, the existence of the double Weyl nodes was rarely mentioned \cite{Cano2017}.  
The double Weyl nodes are realized only when an effective $4 \times 4$ effective ${\mathbf{k}}\cdot{\mathbf{p}}$ model (shown below) 
includes cubic terms in $k$, such as $B({\mathbf k})=-\frac{1}{2}B_3 k_{-}^2k_z$. This $B({\mathbf k})$ term respects the crystal symmetries 
and it couples the $j_z=\pm3/2$ and $j_z=\mp1/2$ bands, where $k_{\pm}$ denote $k_x \pm ik_y$. The Hamiltonian reads
\begin{equation}
H_{4 \times 4} =
\epsilon_0({\mathbf k})+\left(\begin{array}{cccc}
 M({\mathbf k}) & A({\mathbf k}) &  0  & B^{\star}({\mathbf k})  \\
 A^{\star}({\mathbf k}) & -M({\mathbf k}) & B^{\star}({\mathbf k}) & 0 \\
 0  & B({\mathbf k}) & M({\mathbf k}) & -A^{\star}({\mathbf k}) \\
 B({\mathbf k}) & 0 & -A({\mathbf k}) & -M({\mathbf k})
 \end{array} \right),
\end{equation}
where $M({\mathbf k})=M_0+M_1k_z^2+M_2(k_{x}^2+k_y^2)$, $A({\mathbf k})=ik_{+}A_0(1+A_1k_z^2+A_2(k_{x}^2+k_{y}^2))$, and
$\epsilon_0({\mathbf k})=C_0+C_1k_z^2+C_2(k_x^2+k_y^2)$.  The parameter values except for $B_3$, $A_1$, and $A_2$ are found in Ref.~\onlinecite{Wang2012}. Note that the above Hamiltonian is in our global coordinates, where our $x$ and $y$ coordinates are reversed 
from those in Ref.~\onlinecite{Wang2012}. Since the cubic terms [$B({\mathbf k})$, $iA_0A_1k_+k_z^2$, and $iA_0A_2k_+(k_{x}^2+k_{y}^2)]$ 
do not affect the linear dispersion near the Dirac nodes in zero ${\mathbf B}$ field, the existence of the terms
cannot be shown from the fitting to the DFT-calculated bands in previous studies \cite{Wang2012}.  Our finding of the double Weyl
nodes is the first direct evidence of the existence of the $B({\mathbf k})$ term.

\subsection{Nodal Rings Formed via Magnetic Field}

\begin{figure*}[htb]
\begin{center}
\includegraphics[width=0.7 \textwidth]{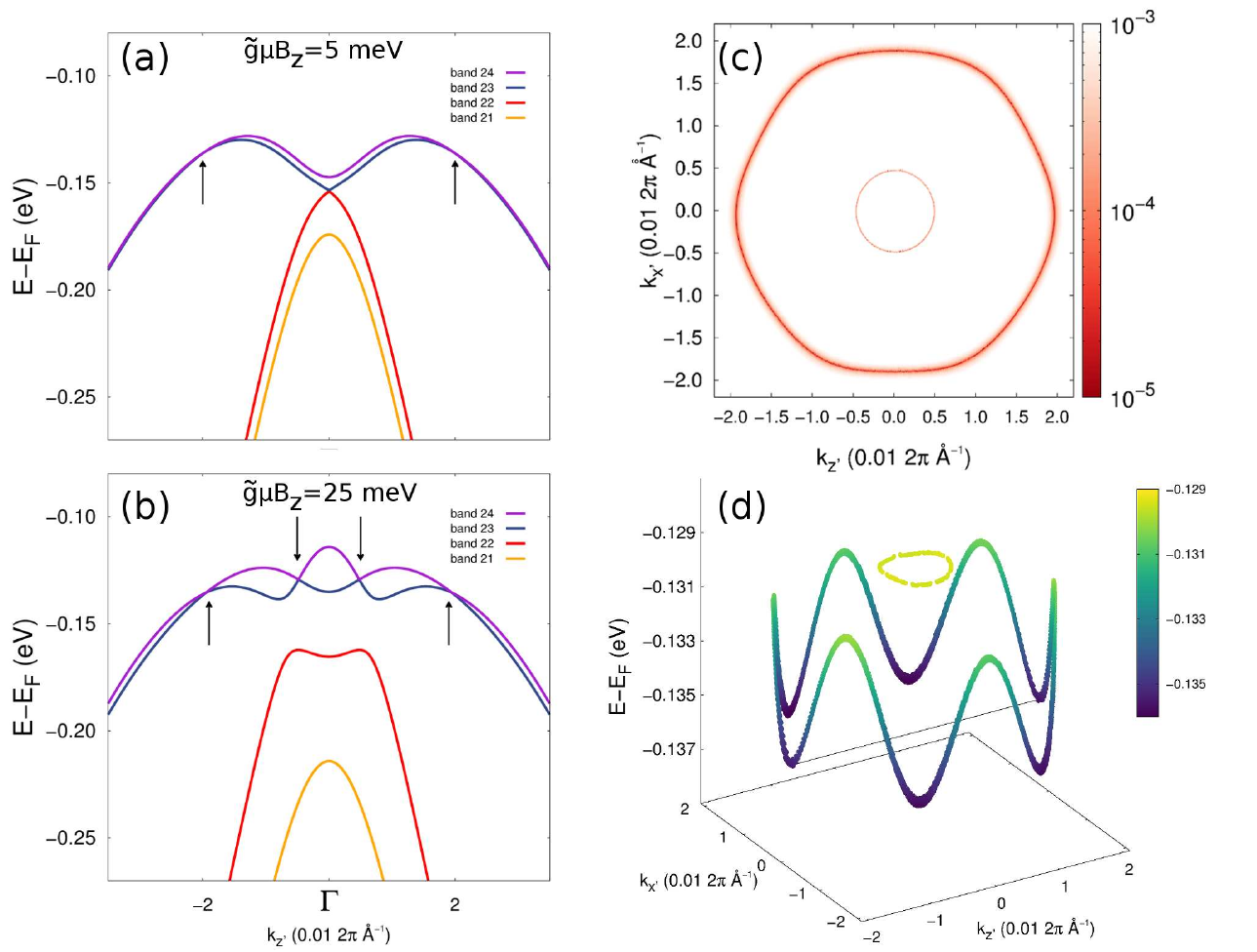}
\caption[Nodal lines]{(Color online) (a) Dispersion of bands $21-24$ of Na$_3$Bi along the $k_{z^{\prime}}$ axis with $\tilde{g}\mu_B B_z$$=$5 meV.  The gapless points are emphasized with arrows. (b) Likewise when $\tilde{g}\mu_B B_z$$=$25 meV.  (c) Nodal rings formed by bands 23 and 24 in the $k_x-k_y$ plane for $\tilde{g}\mu_B B_z$$=$25~meV.  The colorscale is a logscale for the size of the bandgap in eV.  (d) Gapless nodal rings in the $k_x-k_y$ plane vs energy for $\tilde{g}\mu_B B_z$$=$25~meV.  The color scale is for the energy of the bands.}
\label{fig:NL}
\end{center}
\end{figure*}

In addition to the four Weyl nodes, we also observe nodal rings in the horizontal mirror plane upon ${\mathbf B}$ field along the $c$ axis,
which is consistent with the result obtained from the effective $4 \times 4$ model \cite{Cano2017}.
Figure~\ref{fig:NL}(a) shows bands $n$$=$$21-24$ along the $k_x$ (or $k_z^{\prime}$) axis.  For $\tilde{g}\mu_B B_z$$=$5~meV, bands 23 and 24 meet
at two points along the $k_x$ axis, while for $\tilde{g}\mu_B B_z$$=$25~meV, the two bands meet at four points. In the $k_x-k_y$ plane
bands 23 and 24 form one nodal ring at low fields like $\tilde{g}\mu_B B_z$$=$5~meV but two nodal rings at higher fields like
$\tilde{g}\mu_B B_z$$=$25~meV, as shown in Fig.~\ref{fig:NL}(a) and (b).  Further, we calculate a $\pi$ Berry phase on a loop-path which interlocks with one nodal ring (one path for each ring), and this demonstrates the protection of the nodal rings due to the mirror symmetry \cite{CHAN2016,Vander1993}.  Not all the gapless points at the nodal rings have the same energy.  As shown in Fig.~\ref{fig:NL}(c) for 
$\tilde{g}\mu_B B_z$$=$25~meV, the inner ring
is more or less at the same energy, while the outer ring has some variations in the energy. This is
not surprising. It has been shown in other nodal ring or line semimetals \cite{XU2017,QUAN2017}. A nodal ring was found in fcc bulk Fe with SOC \cite{GM_Vander}. In most reported nodal ring or line semimetals, the nodal rings or lines become gapped except for a few points in the presence
of SOC. However, the nodal rings persist with SOC in TRS-broken WSMs.

\subsection{Evolution of Fermi Arcs as a Function of Chemical Potential}

%Here we examine the Fermi arcs and other surface states which arise from the magnetically-induced single \emph{and} double Weyl nodes.
% observation (result) first and then analysis.

\begin{figure*}[!]
\begin{center}
\includegraphics[width=0.95 \textwidth]{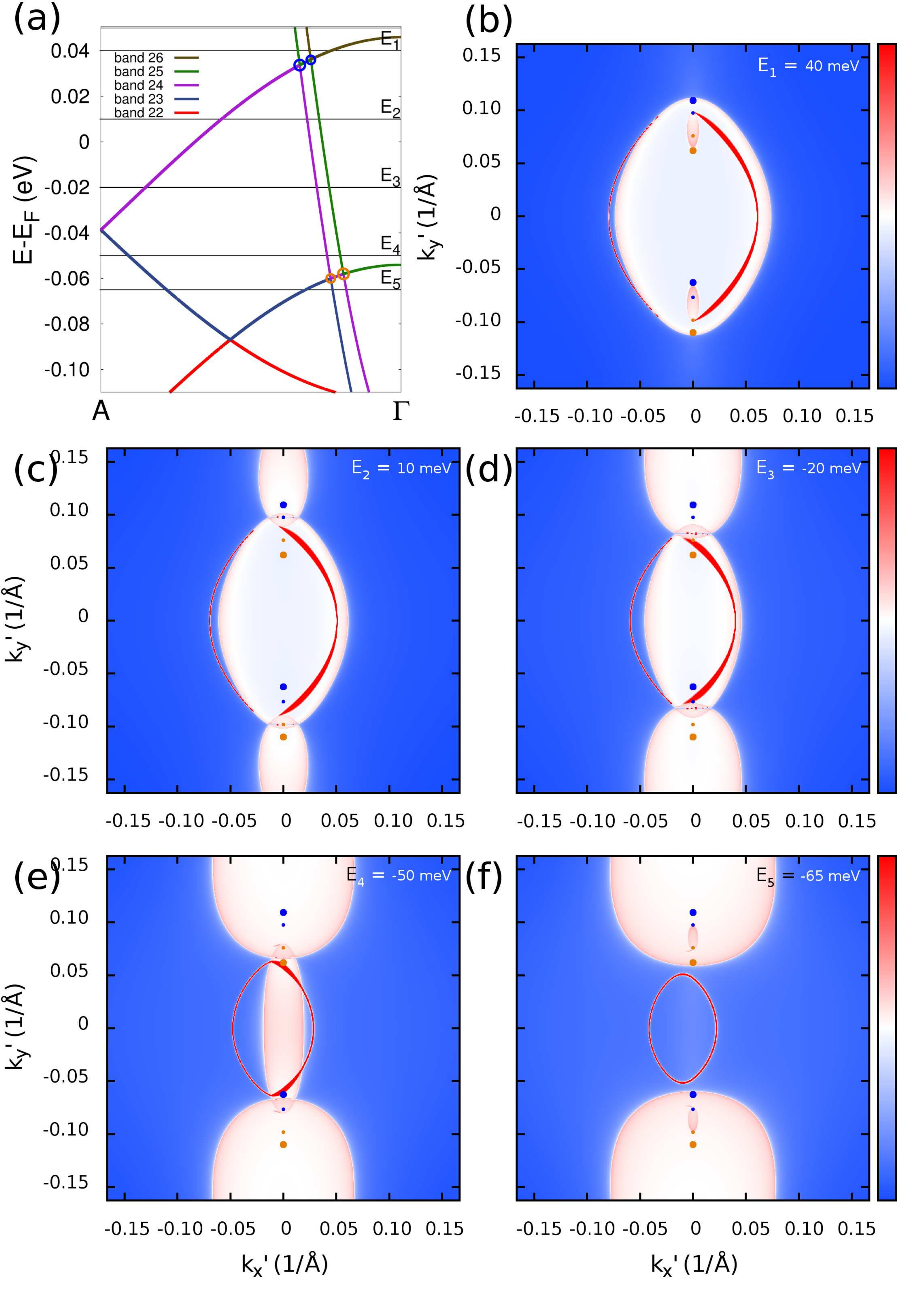}
\caption[Fermi arcs and surface states in a wider view of the Brillouin zone]{(Caption on page \pageref{pagecap1}.)}
\label{fig:FASSBZ}
\end{center}
\end{figure*}

\begin{figure*}[!]
\begin{center}
\includegraphics[width=0.95 \textwidth]{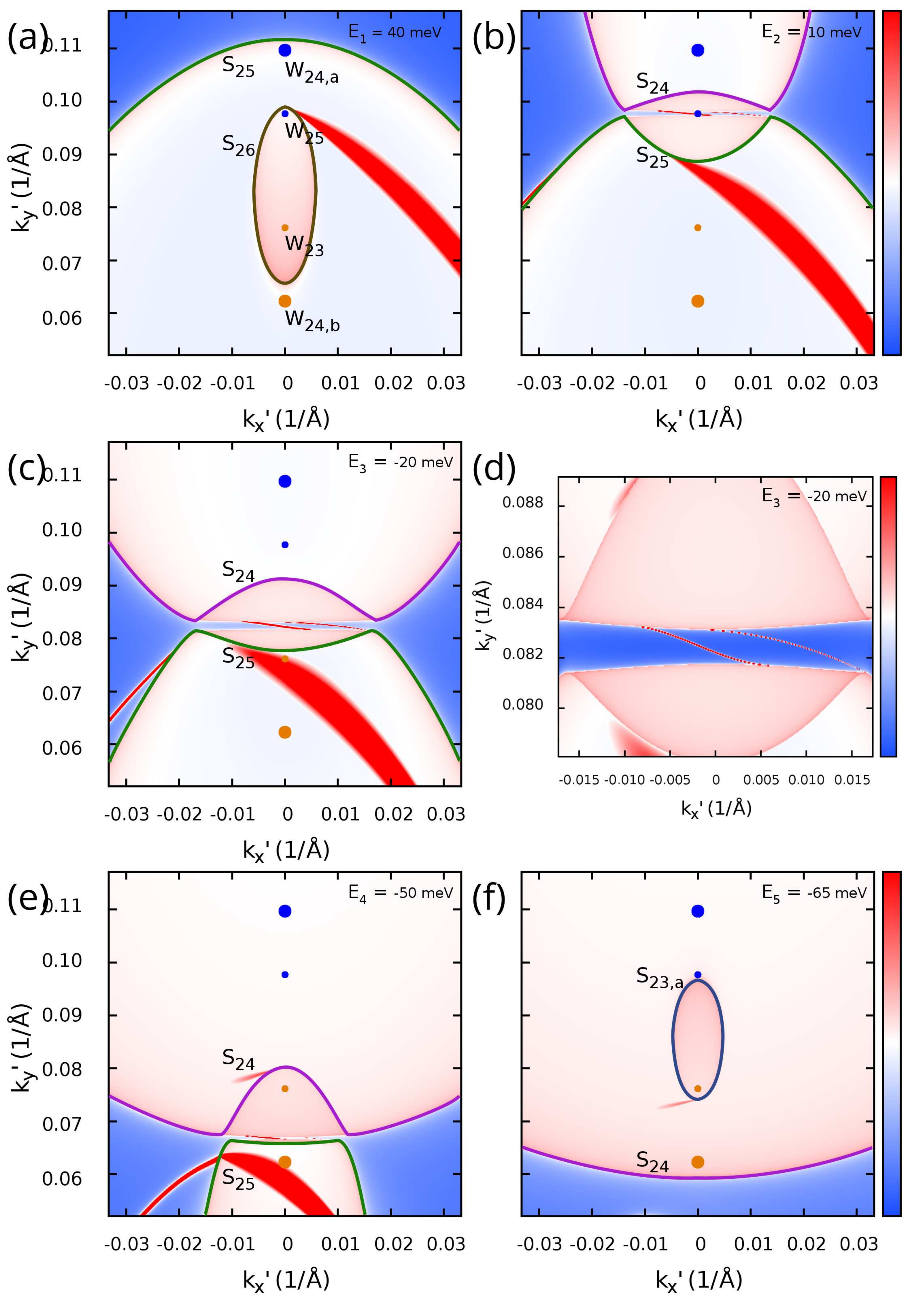}
\caption[Fermi arcs in an applied magnetic field]{(Caption on page \pageref{pagecap2}.)}
\label{fig:FASSzoom}
\end{center}
\end{figure*}

We now present our calculation of Fermi arc surface states at a single surface as a function of chemical potential $\mu$ in the presence of 
${\mathbf B}$ field.  We compute the bulk states and the surface states for a semi-infinite slab with the (1${\bar 2}$0) surface by using 
an iterative Green's
function method \cite{MPLSancho} as implemented in {\sc Wannier Tools} \cite{WTOOLS}.  We use three principle layers in the iterative method.  Figure~\ref{fig:FASSBZ}(a) shows a zoom-in of the five bands $n$$=$$22-26$ near the Fermi level, colored in correspondence with Fig~\ref{fig:NL} and for future reference.
Figure~\ref{fig:FASSBZ}(b)-(f) shows the calculated Fermi arcs at five values of chemical potential, $E_1$,...,$E_5$ as indicated in Fig~\ref{fig:FASSBZ}(a).  The Fermi arcs are indicated as red curves, while the bulk states are shown as white and pale
red.  The single and double Weyl points are marked as small and large circles, respectively.  The orange (blue) color is for the positive (negative) chirality.    Weyl nodes on the $+k_{y^{\prime}}$ axis are related to the Weyl nodes on the $-k_{y^{\prime}}$ axis by the horizontal mirror plane.  Figure~\ref{fig:FASSzoom} shows zoom-ins
of the Fermi arcs and bulk states near the Weyl points at the five energies.  We denote the boundary of a Fermi surface (FS) volume ${\cal V}$
of band $n$ at different $\mu$ values as $S_n$.  The color of the boundary in Fig.~\ref{fig:FASSzoom} corresponds to the denoted band index as in Fig.~\ref{fig:FASSBZ}(a).

At $\mu$$=$$E_1$$=$$40$~meV  we find two Fermi arc surface states in Fig.~\ref{fig:FASSBZ}(b).
One of them terminates tangentially on each of two FS sheets such as $S_{26}$ (labeled in Fig.~\ref{fig:FASSzoom}) and its mirror partner,
whereas the other Fermi arc terminates tangentially
on FS sheet $S_{25}$, as shown in Figs.~\ref{fig:FASSBZ}(b) and ~\ref{fig:FASSzoom}(a).  At both $\mu$$=$$E_2$$=$$10$~meV  and
$\mu$$=$$E_3$$=$$-20$~meV , we observe four Fermi arc surface states in the half-BZ; two arcs end tangentially on FS sheet $S_{25}$ and
the other two (appearing within the gap between $S_{24}$ and $S_{25}$) terminate tangentially on $S_{24}$, as shown in Figs.~\ref{fig:FASSBZ}(c) and (d) and~\ref{fig:FASSzoom}(b) and (c).  Figure~\ref{fig:FASSzoom}(d) provides a zoom-in view of the latter two Fermi arcs, and clearly shows the
gap between the two different FS sheets.  This zoom-in view is qualitatively the same for
$\mu$$=$$E_2$, $E_3$ and $E_4$, discussed next.  At $\mu$$=$$E_4$$=$$-50$~meV we find one surface-state loop connected through $S_{25}$ and three
Fermi arcs ending tangentially on $S_{24}$ [Fig.~\ref{fig:FASSzoom}(e) and Fig.~\ref{fig:FASSBZ}(e)].  In this case, the closed surface state is
similar to a topological-insulator-like surface state in external in-plane ${\mathbf B}$ fields \cite{ZYUZ11,PERSH12}, and it terminates
{\it non-tangentially} on $S_{25}$.  The two Fermi arcs within the gap between $S_{24}$ and $S_{25}$ look similar to
the cases of $E_2$ and $E_3$.  The third Fermi arc looks like a short whisker which forms off of the ``crumpled-nosecone" of
$S_{24}$ and is lost into the projected bulk states.  At $\mu$$=$$E_5$$=$$-65$~meV we observe one topological-insulator-like
closed surface state well separated from the bulk states and one whisker-like Fermi arc terminating tangentially on $S_{23,a}$
[Fig.~\ref{fig:FASSzoom}(f) and Fig.~\ref{fig:FASSBZ}(f)].

\addtocounter{figure}{-2}
\begin{figure}[htb!]
\caption[]{(Color online) (a) Zoom-in on the band structure showing the five choices of chemical potential.  (b-f) Fermi arcs and other surface states on the $(1\bar{2}0)$ in an area one quarter of the full BZ area.  The panels show $E_1-E_5$ respectively, which are +40, +10, -20, -50 and -65 meV.  The colorscale is a chimera of the spectral densities of the bulk and surface Green's functions: significant weight of the bulk Green's function appears white or light red, and weight on the Fermi arc surface states appears as strictly the darkest red.  This makes it apparent that many of the surface states attach tangentially to the bulk Fermi surface, and at low enough energy the surface states are disconnected from the bulk Fermi surface.}
\label{pagecap1}
\end{figure}

\begin{figure}[htb!]
\caption[]{(Color online) (a-f) Fermi arcs on the $(1\bar{2}0)$ surface, magnified near to the Weyl nodes at energies $E_1-E_5$ respectively, which are +40, +10, -20, -50 and -65 meV. (d) is a zoom-in on the two Fermi arc states at $E_3$$=$-20 meV connecting the $C_{24}$$=$+2 and $C_{25}$$=$0 Fermi surface sheets. The colorscale in all panels is a chimera of the spectral densities of the bulk and surface Green's functions: significant weight of the bulk Green's function appears white or light red, and weight on the Fermi arc surface states appears as strictly the darkest red.  This makes it apparent that many of the surface states attach tangentially to the bulk Fermi surface.}
\label{pagecap2}
\end{figure}

\subsection{Analysis of Fermi-Arcs Evolution from Fermi Surface Chern numbers}

In order to understand the evolution of the Fermi arcs as a function of $\mu$, we examine the Chern numbers of disjoint FS sheets at different
$\mu$ values around the Weyl nodes following Gos\'albez-Mart\'inez {\it et al.} \cite{GM_Vander}.  The Chern number, $C_n$, of each FS sheet
$S_{n,\alpha}$ enclosing a volume ${\cal V}$ arising from band $n$ is given by
\begin{equation}
C_n = \sum_{W_{n,\alpha}\in {\cal V}} \chi_{n,\alpha} - \sum_{W_{n-1,\alpha}\in {\cal V}} \chi_{n-1,\alpha},
\label{eqn:chernFS}
\end{equation}
where $\chi_{n,\alpha}$ is the chirality of the $\alpha^{th}$ Weyl node connecting bands $n$ and $n+1$, $W_{n,\alpha}$.
The sum is over all Weyl nodes interior to the FS sheet; the outward pointing normal vector points toward the exterior of the FS for electron-pockets and the reverse for hole-pockets.  With this convention, the band crossing point becomes a source of Berry curvature flux in the lower band ($n$)
and a sink of Berry curvature flux in the upper band ($n+1$). Table \ref{tab:FSChern} lists the calculated FS Chern numbers of each Fermi sheet belonging to each band at each of the five chemical potential values.

\setlength{\tabcolsep}{14pt}
\begin{table*}[htb]
\centering
\caption[Fermi surface Chern numbers]{$C_n$ for each Fermi surface sheet, $S_n$, as a function of chemical potential (meV).
Only Fermi surface sheets appearing in the $z>0$ (or $y^{\prime}>0$) half-BZ are listed. Not all $S_{n,\alpha}$ are
shown in Fig.~\ref{fig:FASSBZ}(a).}
\begin{tabular}{c c c c c c}
\hline\hline
$n,\alpha$ in $S_{n,\alpha}$ & $E_1$$=$$+$40 & $E_2$$=$$+$10 & $E_3$$=$$-$20 & $E_4$$=$$-$50 & $E_5$$=$$-$65 \\ [0.5ex]
\hline
26 & +1 & -  & -  & -  & - \\
25 & 0  & 0  & 0  & 0  & - \\
24 & -  & +2 & +2 & 0 & 0 \\
23,$a$($b$) & -  & -  & -  & 0  & $-$1 (0) \\
\hline
\end{tabular}
\label{tab:FSChern}
\end{table*}

At $\mu$$=$$E_1$, the chemical potential meets bands 25 and 26 and so $C_{25}$ and $C_{26}$ are relevant to our analysis
[Figs.~\ref{fig:FASSBZ}(b) and~\ref{fig:FASSzoom}(a)].  We find that a small electron-pocket ellipsoid $S_{26,a}$ (brown) encloses
$W_{25}$ and $W_{23}$.  Only $W_{25}$ enters into the calculation of $C_{26}$ according to Eq.~(\ref{eqn:chernFS}) and so FS sheet
$S_{26,a}$ inherits a Chern number of $C_{26}=-\chi_{25}=+1$.  The other electron-pocket ellipsoid $S_{25}$ (green) encloses
all eight Weyl nodes  in the full BZ. Among them, all the Weyl nodes except for $W_{23}$ and its mirror-symmetry partner are relevant, and its FS
Chern number is zero, i.e. $C_{25}$$=$$0$.  This is
alternatively understood from the fact that $S_{25}$ encloses the parity-invariant $\Gamma$ point, so its Chern number must be zero.
From the calculations of $C_{25}$ and $C_{26}$, we might expect one Fermi arc per surface terminating on $S_{26}$ and zero terminating on
$S_{25}$.  However, Haldane \cite{HALDANE} points out that Fermi arcs arising from Weyl nodes higher or lower in energy may still be observed
away from the energy of the Weyl node (even when the FS Chern number is zero by enclosing Weyl nodes of opposite chirality), so long as they
terminate tangentially and respect the Chern number of the Fermi surface.  For $C_{25}$$=$$0$, such a state would have to originate from and
terminate on the same surface.  This analysis agrees with the observed two Fermi arcs discussed in Sec.~IV.D [Fig.~\ref{fig:FASSBZ}(a)].

\begin{figure}[h!]
\begin{center}
\includegraphics[width=0.3 \textwidth]{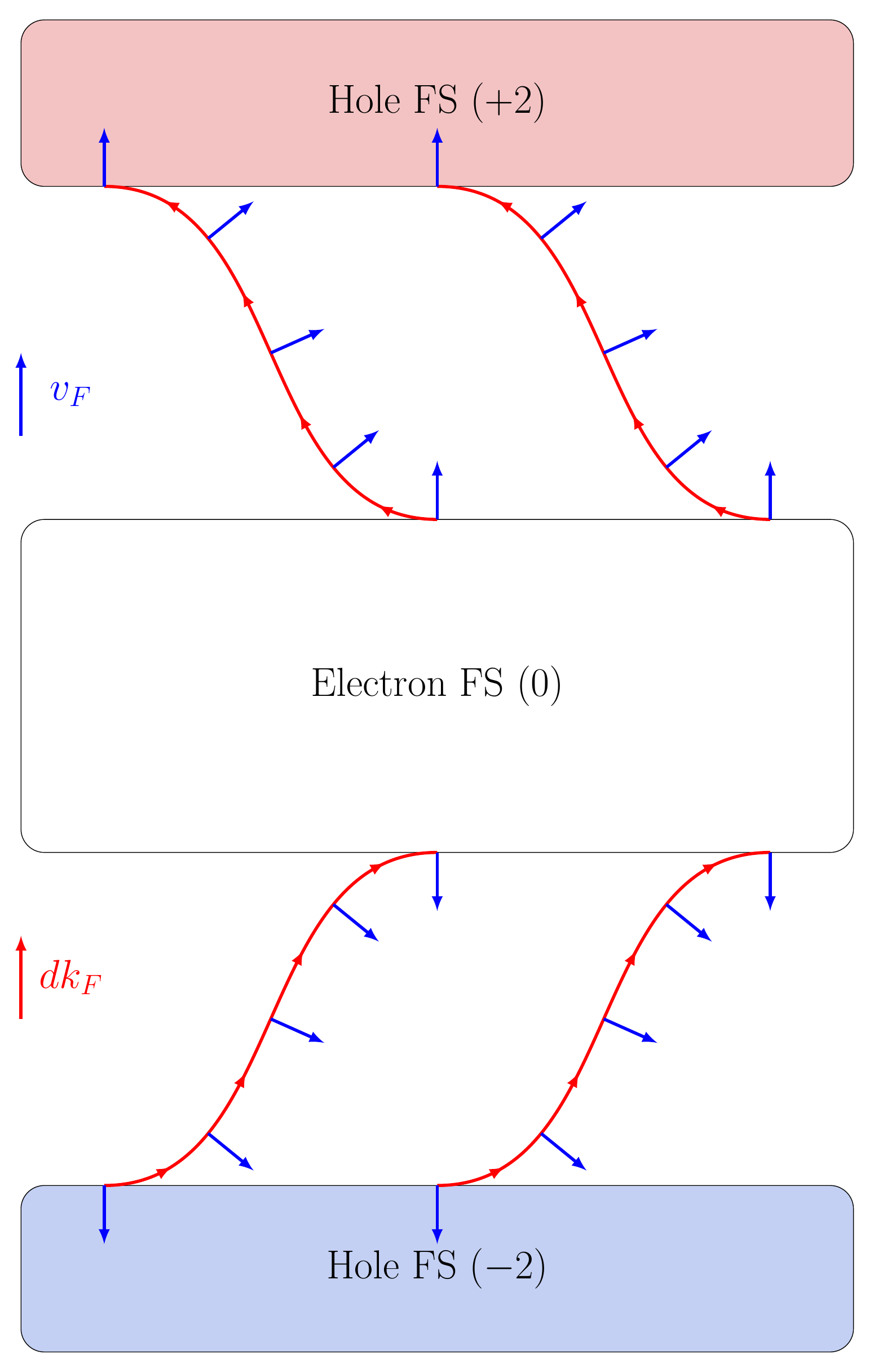}
\caption[attachment of Fermi arcs at E3]{(Color online) Schematic representation of the attachment of Fermi arcs in the full surface Brillouin zone.  The upper-half of this figure should be compared with Fig.~\ref{fig:FASSzoom}(d).  The Fermi arcs on one surface attach tangentially to the Fermi surface (FS) sheets.  The local Fermi velocities, $v_F$, (blue arrows) match at the points of attachment; pointing outward (inward) for electron (hole) pockets.  Choosing a consistent direction for the differential Fermi vector, $dk_F$, (red arrows) along the Fermi arc as in Ref.~\onlinecite{HALDANE} (i.e. $(\hat{n}\times \vec{v}_F)\cdot d\vec{k}_F > 0$), there is a net Fermi arc flux associated with the FS sheets with nontrivial Chern number, while such flux is zero into the FS with zero Chern number (two arcs enter and two arcs exit). }
\label{fig:haldane}
\end{center}
\end{figure}

At $\mu$$=$$E_2$, bands 24 and 25 meet the chemical potential and so $C_{24}$ and $C_{25}$ are relevant to the counting of the Fermi arcs
[Figs.~\ref{fig:FASSzoom}(c) and~\ref{fig:FASSzoom}(b)]. For $S_{25}$ a similar analysis to the case of $\mu$$=$$E_1$ can be applied and thus
$C_{25}$$=$$0$.  The hole-pocket $S_{24}$ (magenta) encloses the double Weyl node $W_{24,a}$, yielding $C_{24}$$=$$+2$.  Note that an extra
minus seems to enter for a hole pocket in Eq.~(\ref{eqn:chernFS}), but only because of the reversed direction of the Fermi velocity vector at the surface of the hole pocket alters the sense of which Weyl nodes are \emph{interior} to the FS sheet.  One may initially expect two Fermi arcs to connect $S_{24}$ and its mirror partner, yet these arcs terminate on $S_{25}$.  Figure~\ref{fig:haldane} represents this case schematically to facilitate discussion.  Though these Fermi arcs in the $k_{y^{\prime}}>0$ half-BZ terminate on $S_{25}$, their mirror symmetric partner states in the $k_{y^{\prime}}<0$ half-BZ do so as well.  In this sense two arcs enter and two arcs exit $S_{25}$, consistent with $C_{25}$$=$$0$.  Meanwhile the separated hole FS sheets, $S_{24}$ and its mirror partner, have a net flow of Fermi arcs (indicated as red arrows in Fig.~\ref{fig:haldane})
into or out of each and in opposite measure, consistent with their nontrivial and opposite FS Chern numbers.
At $\mu$$=$$E_3$, the analysis is much the same as at the preceding energy except that the hole-pocket $S_{24}$ encloses an extra Weyl node
$W_{25}$ [Fig.~\ref{fig:FASSzoom}(c)].  The extra Weyl node connects bands with indices which are irrelevant for the calculation of the
hole-pocket Chern number ($C_{24}$), and so $C_{24}$ does not change.

At $\mu$$=$$E_4$, the chemical potential meets bands 23, 24, and 25, and so $C_{23}$, $C_{24}$, and $C_{25}$ are relevant [Fig.~\ref{fig:FASSzoom}(e)].
%The crossing point between bands 22 and 23 does not carry any chirality and the electron-pocket $S_{23}$ (blue) encloses $W_{23}$ and its partner
%with opposite chirality across the $\Gamma$ point.
The crossing point between bands 23 and 24 at the BZ boundary (A point) does not carry any chirality and the hole-pocket $S_{23}$ (not shown)
encloses this crossing point.  Thus we obtain $C_{23}$$=$$0$.
%A similar analysis is applied to $S_{25}$ and so $C_{25}=0$.
The electron-pocket $S_{25}$ encloses $W_{24,b}$ and its partner with opposite chirality across the $\Gamma$ point, and so $C_{25}$$=$$0$.
Now $S_{24}$ is roughly ellipsoidal except for a ``crumpled-nosecone" shape near where it avoids $S_{25}$.  The Weyl node $W_{23}$ is actually \emph{exterior to} $S_{24}$.  $S_{24}$ extends all the way across the BZ boundary (the A point) enclosing $W_{24,a}$ and its partner with
opposite chirality, so we have $C_{24}$$=$$0$.  Our analysis shows that the number of Fermi arcs is not constrained at this chemical potential.

At $\mu$$=$$E_5$, bands 23 and 24 cross the chemical potential which is below the energies of all four Weyl nodes.  In the half-BZ ($y^{\prime}>0$),
there are two disjoint hole-pockets $S_{23,a}$ (blue) and $S_{23,b}$ (not shown in Fig.~\ref{fig:FASSzoom}(f)). The former hole pocket encloses only $W_{23}$ while the latter pocket encloses the A point, so $C_{23,a(b)}$$=$$-1 (0)$. The hole-pocket $S_{24}$ (magenta) reaches all the way across the the BZ boundary, enclosing the parity-invariant point A (not shown) and also all of the Weyl nodes and their partners.  Thus $C_{24}$$=$$0$.  This analysis dictates one Fermi arc tangentially terminating on $S_{23,a}$, which corroborates our result.

\section{Conclusion}

In summary, we have developed a WF-TB model for the topological DSM, Na$_3$Bi, which reproduces the DFT-calculated band structure well while retaining the symmetries of the crystal.  The projected atomic Wannier functions are atom-centered with larger spread than maximally localized WFs.  Atomic-like SOC was included, and we investigated the formation of line nodes in the mirror plane and splitting of the Dirac nodes into multiple Weyl nodes in an applied magnetic field.  We found that each Dirac node splits into pairs of Weyl nodes with chiral charges $\pm 1$ and $\pm 2$ from the calculations of Berry curvature.  By carefully considering the Chern number of associated Fermi surface sheets, we detailed the interesting development of Fermi arc and other topological surface states as a function of chemical potential consistent with the topological charges of the Weyl nodes.  Our tight-binding model can be used to calculate novel properties induced by the nonzero Berry curvature, and its qualitative features can be applied in another experimentally observed topological DSM, Cd$_3$As$_2$.

\begin{acknowledgments}
J.V. was supported by the National Science Foundation (NSF) CREST Center for Interface Design and Engineered Assembly of Low-dimensional Systems (IDEALS) under NSF grant number HRD1547830. The computational support was provided by San Diego Supercomputer Center (SDSC) under DMR060009N and VT Advanced Research Computing (ARC).
\end{acknowledgments}

\appendix
\section{Electronic Structures Using the Non-primitive Unit Cell}

\begin{figure*}[]
\begin{center}
\includegraphics[width=0.4\textwidth]{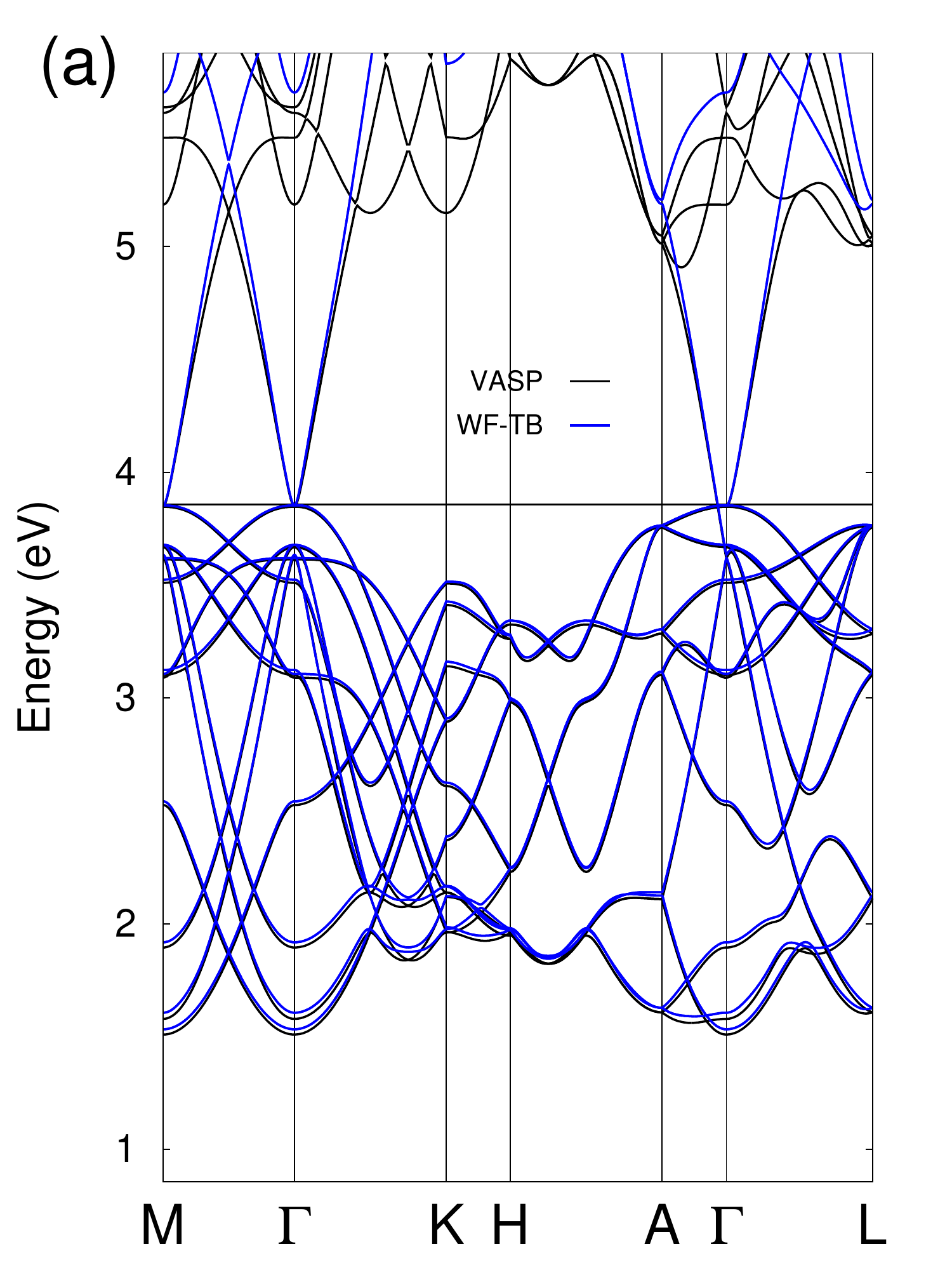}
\includegraphics[width=0.4\textwidth]{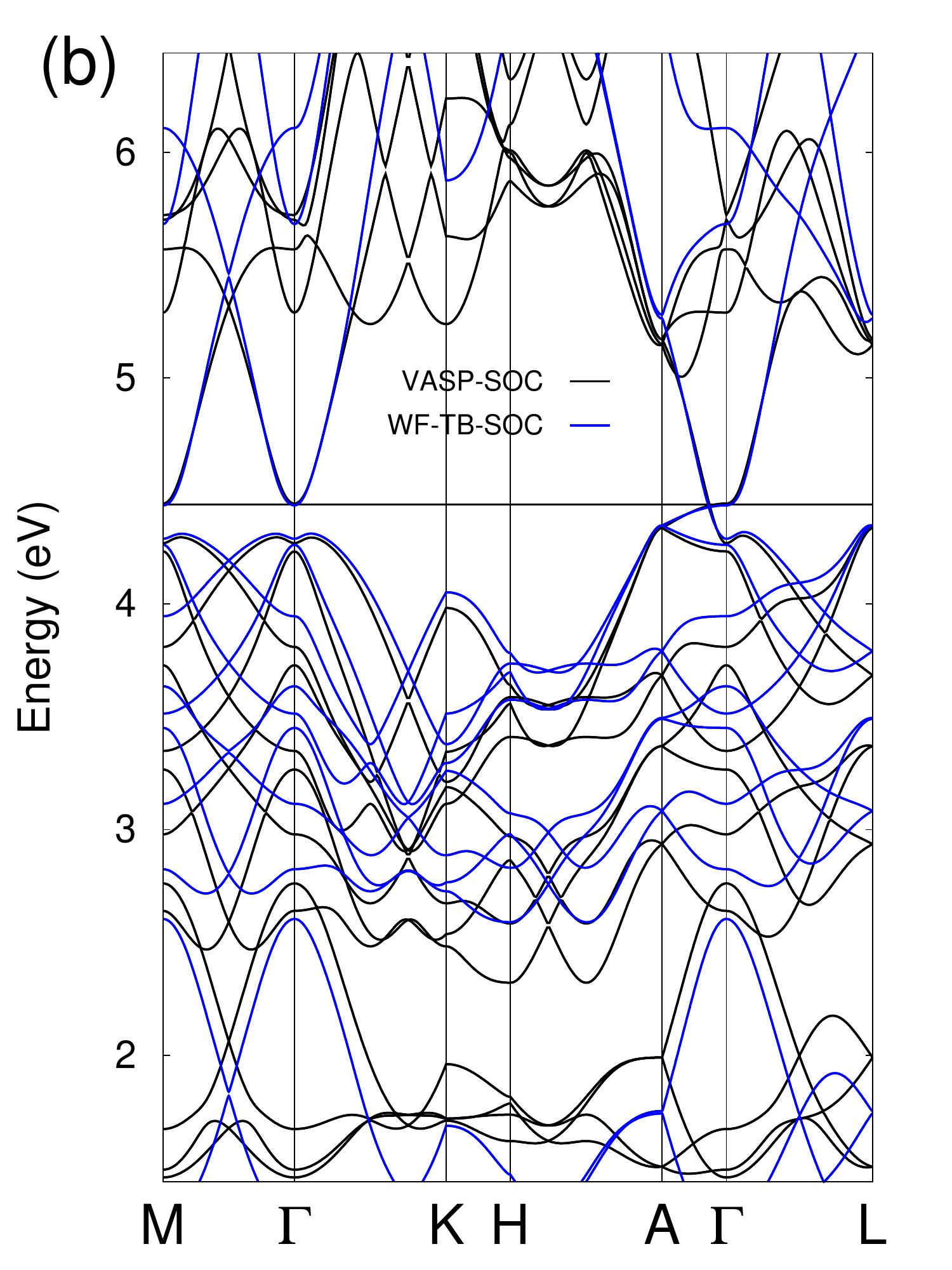}
\caption[nonprimitive unit cell band overlap]{(Color online) Band structures of bulk Na$_3$Bi in the non-primitive unit cell (a) without SOC and (b) with SOC.  The black (dark) bands show the the DFT-calculated bands using VASP, and the blue (light) bands are calculated from the WF-TB model.  The high-symmetry point labels correspond to those in the primitive unit cell, so the band structures clearly show band-folding, e.g. midway along M$\Gamma$ where the non-primitive unit cell BZ ends.}
\label{fig:appfig}
\end{center}
\end{figure*}

In Fig. \ref{fig:bulk_120} of the main text, we compared the band structure of the WF-TB model in the non-primitive unit cell to the familiar band structure of bulk Na$_3$Bi calculated in the primitive unit cell.  In this section, we present the band structure of Na$_3$Bi calculated using VASP versus the band structure calculated using the WF-TB model in the non-primitive unit cell for ease of comparison, given the band-folding.    Figure \ref{fig:appfig}(a) shows the bands in the non-primitive unit cell without SOC; there is excellent agreement in the whole of the valence band and up to even 1 eV above the Fermi level.  Figure \ref{fig:appfig}(b) shows the bands with SOC included; while far from the Fermi level the bands are identical in dispersion but differ by a vertical shift in energy, there is excellent agreement within [-0.5,0.5] eV around the Fermi level.  Hence our WF-TB model reproduces the DFT electronic structure very well with and without SOC even for the non-primitive unit cell.

%\newpage
%\clearpage
%\FloatBarrier


\begin{thebibliography}{99}

%=============INTRODUCTION===========%
\bibitem{Nielsen1981}
H. B. Nielsen and M. Ninomiya, Phys. Lett. B {\bf 105}, 219-223 (1981).

\bibitem{XWan2011}
X. Wan, A. M. Turner, A. Vishwanath, and S. Y. Savrasov, Phys. Rev. B {\bf 83}, 205101 (2011).

\bibitem{Xu2011}
G. Xu, H. Weng, Z. Wang, X. Dai, and Z. Fang, Phys. Rev. Lett. {\bf 107}, 186806 (2011).

\bibitem{ChenFang2012}
C. Fang, M. J. Gilbert, X. Dai, and B. A. Bernevig, Phys. Rev. Lett. {\bf 108}, 266802 (2012).
%Multi-Weyl Topological Semimetals Stabilized by Point Group Symmetry

%TaAs theory
\bibitem{HWeng2015}
H. Weng, C. Fang, Z. Fang, B. A. Bernevig, and X. Dai, Phys. Rev. X {\bf 5}, 011029 (2015).

%TaAs exp
\bibitem{BQLv2015}
%B. Q. Lv {\it et. al.}
B. Q. Lv, H. M. Weng, B. B. Fu, X. P. Wang, H. Miao, J. Ma, P. Richard, X. C. Huang, L. X. Zhao, G. F. Chen, Z. Fang, X. Dai, T. Qian, and H. Ding, Phys. Rev. X {\bf 5}, 031013 (2015).

\bibitem{LXYang2015}
L. X. Yang, Z. K. Liu, Y. Sun, H. Peng, H. F. Yang, T. Zhang, B. Zhou, Y. Zhang, Y. F. Guo, M. Rahn, D. Prabhakaran, Z. Hussain, S.-K. Mo, C. Felser, B. Yan, and Y. L. Chen, Nat. Phys. {\bf 11}, 728-732 (2015).

\bibitem{Bernevig2012}
%multi-weyl topological semimetals
C. Fang, M. J. Gilbert, X. Dai, and B. A. Bernevig, Phys. Rev. Lett. {\bf 108}, 266802 (2012).

\bibitem{SOLU15}
%type-2 wsm
A. A. Soluyanov, D. Gresch, Z. Wang, Q. S. Wu, M. Troyer, X. Dai, and B. A. Bernevig, Nature {\bf 527}, 495-498 (2015).

\bibitem{Li2017}
P. Li, Y. Wen, X. He, Q. Zhang, C. Xia, Z.-M. Yu, S. A. Yang, Z. Zhu, H. N. Alshareef, and X.-X. Zhang, Nat. Commun. {\bf 8}, 2150 (2017).

%Chiral magnetic effect/ chiral anomaly
\bibitem{SON2012}
D. T. Son and N. Yamamoto, Phys. Rev. Lett. {\bf 109}, 181602 (2012).

\bibitem{SON2013}
D. T. Son and B. Z. Spivak, Phys. Rev. B {\bf 88}, 104412 (2013).

\bibitem{Burkov2015}
A. A. Burkov, J. Phys.: Condens. Matter {\bf 27}, 113201 (2015).

\bibitem{SHARMA16}
%Nernst and magnetothemral conductivity in a lattice model of Weyl fermions
G. Sharma, P. Goswami, and S. Tewari, Phys. Rev. B {\bf 93}, 035116 (2016).

\bibitem{SHARMA17}
%Nernst effect in Dirac and inversion-asymmetric Weyl semimetals
G. Sharma, C. Moore, S. Saha, and S. Tewari, Phys. Rev. B {\bf 96}, 195119 (2017).

%Quantum oscillations from surface Fermi arcs in Weyl and Dirac semimetals.
\bibitem{POTT14}
A. C. Potter, I. Kimchi, and A. Vishwanath, Nat. Commun. {\bf 5}, 5161 (2014).
\bibitem{MOLL16}
P. J. W. Moll, N. L. Nair, T. Helm, A. C. Potter, I. Kimchi, A. Vishwanath, and J. G. Analytis, Nature {\bf 535}, 266-270 (2016).

%AFM TRS-broken Weyl
\begin{comment}
\bibitem{KIYO2016}
N. Kiyohara, T. Tomita, and S. Nakatsuji, Phys. Rev. Applied {\bf 5}, 064009 (2016).
%Mn3Ge too
\bibitem{YANG2017}
H. Yang, Y. Sun, Y. Zhang, W.-J. Shi, S. S. P. Parkin, and B. Yan, New J. Phys. {\bf 19}, 015008 (2017).
%Weyl in chiral afm materials Mn3Ge and Mn3Sn
\bibitem{LIU2017}
J. Liu and L. Balents, Phys. Rev. Lett. {\bf 119}, 087202 (2017).
% more on Mn3Sn/Ge
\end{comment}
\bibitem{HUANG2017}
S. Huang, J. Kim, W. A. Shelton, E. W. Plummer, and R. Jin, Proc. Natl. Acad. Sci. USA {\bf 114}, 6256-6261 (2017).
%BaMnSb2
%

\bibitem{Ando2017}
M. Sato and Y. Ando, Rep. Prog. Phys. {\bf 80}, 076501 (2017).

\bibitem{BJYang2014}
B.-J. Yang and N. Nagaosa, Nat. Commun. {\bf 5}, 4898 (2014).
%Yang, B.-J. and Nagaosa, N.
%{\it Nature Comm.} {\bf 5,} 4898 (2014).
%DOI: 10.1038/ncomms5898


%=========Experiments on Na3Bi and Cd3As2==========%
\bibitem{LIU14_CdAs}
Z. K. Liu, J. Jiang, B. Zhou, Z. J. Wang, Y. Zhang, H. M. Weng, D. Prabhakaran, S.-K. Mo, H. Peng, P. Dudin, T. Kim, M. Hoesch, Z. Fang, X. Dai, Z. X. Shen, D. L. Feng, Z. Hussain, and Y. L. Chen, Nat. Mater. {\bf 13}, 677-681 (2014).
%A stable three-dimensional topological Dirac semimetal Cd$_3$As$_2$.

\bibitem{LIU14}
Z. K. Liu, B. Zhou, Y. Zhang, Z. J. Wang, H. M. Weng, D. Prabhakaran, S.-K. Mo, Z. X. Shen, Z. Fang, X. Dai, Z. Hussain, Y. L. Chen, Science {\bf 343}, 864-867 (2014).
%Discovery of a three-dimensional topological Dirac semimetal, Na$_3$Bi.

\bibitem{Wang2012}
%PRB2012
Z. Wang, Y. Sun, X.-Q. Chen, C. Franchini, G. Xu, H. Weng, X. Dai, and Z. Fang, Phys. Rev. B {\bf 85}, 195320 (2012).

\bibitem{WANG13_CdAs}
Z. Wang, H. Weng, Q. Wu, X. Dai, and Z. Fang, Phys. Rev. B {\bf 88}, 125427 (2013).
%Three-dimensional Dirac semimetal and quantum transport in Cd$_3$As$_2$.

\bibitem{Gorbar2015}
E. V. Gorbar, V. A. Miransky, I. A. Shovkovy, and P. O. Sukhachov, Phys. Rev. B {\bf 91}, 235138 (2015).
%Surface Fermi arcs in $Z_2$ Weyl semimetals A$_3$Bi (A=Na,K,Rb).
%DOI: 10.1103/PhysRevB.91.235138

\bibitem{Cano2017}
%chiral anomaly factory
J. Cano, B. Bradlyn, Z. Wang, M. Hirschberger, N. P. Ong, and B. A. Bernevig, Phys. Rev. B {\bf 95}, 161306(R) (2017).

\bibitem{Soluyanov2011}
A. A. Soluyanov and D. Vanderbilt, Phys. Rev. B {\bf 83}, 035108 (2011).

\bibitem{Thonhauser2006}
T. Thonhauser and D. Vanderbilt, Phys. Rev. B {\bf 74}, 235111 (2006).

\bibitem{WZhang2010}
W. Zhang, R. Yu, H.-J. Zhang, X. Dai, and Z. Fang, New J. Phys. {\bf 12} 065013 (2010).

\bibitem{Yu2015}
R. Yu, H. Weng, Z. Fang, X. Dai, and X. Hu, Phys. Rev. Lett. {\bf 115}, 036807 (2015).

%==================END INTRO===============%

%=============Geometry, Symmetry, Methods==========%
\bibitem{JV2}
%Cite self for Geometry Figure
J. W. Villanova, E. Barnes, and K. Park, Nano Lett. {\bf 17}, 963-972 (2017).

\bibitem{VASP} 
G. Kresse and D. Joubert, Phys. Rev. B {\bf 59}, 1758 (1999).

\bibitem{QE}
P. Giannozzi {\it et al}, J. of Physics: Condens. Matt. {\bf 29}, 465901 (2017).
%O. Andreussi, T. Brumme, O. Bunau, M. B. Nardelli, M. Calandra, R. Car, C. Cavazzoni, D. Ceresoli, M. Cococcioni, N. Colonna, I. Carnimeo, A. Dal Corso, S. de Gironcoli, P. Delugas, R. A. DiStasio Jr, A. Ferretti, A. Floris, G. Fratesi, G. Fugallo, R. Gebauer, U. Gerstmann, F. Giustino, T. Gorni, J. Jia, M. Kawamura, H.-Y. Ko, A. Kokalj, E. K\"u\c{c}\"ukbenli, M. Lazzeri, M. Marsili, N. Marzari, F. Mauri, N. L. Nguyen, H.-V. Nguyen, A. Otero-de-la-Roza, L. Paulatto, S. Ponc\'e, D. Rocca, R. Sabatini, B. Santra, M. Schlipf, A. P. Seitsonen, A. Smogunov, I. Timrov, T. Thonhauser, P. Umari, N. Vast, X. Wu, and S. Baroni

\bibitem{Wannier90}
A. A. Mostofi, J. R. Yates, G. Pizzi, Y. S. Lee, I. Souza, D. Vanderbilt, and N. Marzari, Comput. Phys. Commun. {\bf 185}, 2309 (2014).

\bibitem{PAW} 
P. E. Bl\"{o}chl, Phys. Rev. B {\bf 50}, 17953 (1994).
\bibitem{PBE}
J. P. Perdew, K. Burke, and M. Ernzerhof, Phys. Rev. Lett. {\bf 77}, 3865 (1996).

\bibitem{PSLIB}
E. K\"u\c{c}\"ukbenli, M. Monni, B. I. Adetunji, X. Ge, G. A. Adebayo, N. Marzari, S. de Gironcoli, and A. Dal Corso, arXiv:1404.3015 (2014).

\bibitem{WannierOG}
G. H. Wannier, Phys. Rev. {\bf 52}, 191 (1937).

\bibitem{Marzari1997}
N. Marzari and D. Vanderbilt, Phys. Rev. B {\bf 56}, 12847 (1997).

\bibitem{Souza2001}
I. Souza, N. Marzari, and D. Vanderbilt, Phys. Rev. B {\bf65}, 035109 (2001).

%%%% topological obstruction %%%%%%%%%%%%
\bibitem{Thouless1984}
D. J. Thouless, J. Phys. C {\bf 17}, L325 (1984).
%%%%%%%%%%%%%%%%%%%%%%%%%%%%%

\bibitem{Mustafa2016}
J. I. Mustafa, S. Coh, M. L. Cohen, and S. G. Louie, Phys. Rev. B {\bf 94}, 125151 (2016).

\bibitem{ZHANG_NJP}
W. Zhang, R. Yu, H.-J. Zhang, X. Dai, and Z. Fang, New J. Phys. {\bf 12} 065013 (2010).

%%G-FACTORS%%
\bibitem{XION15}
J. Xiong, S. K. Kushwaha, T. Liang, J. W. Krizan, M. Hirschberger, W. Wang, R. J. Cava, and N. P. Ong, Science {\bf 350}, 413-416 (2015).
%Evidence for the chiral anomaly in the Dirac semimetal Na$_3$Bi.
%DOI: 10.1126/science.aac6089

\bibitem{JEON14}
S. Jeon, B. B. Zhou, A. Gyenis, B. E. Feldman, I. Kimchi, A. C. Potter, Q. D. Gibson, R. J. Cava, A. Vishwanath, and A. Yazdani, Nat. Mat. {\bf 13}, 851-856 (2014).
%%END G-FACTORS%%

\bibitem{WTOOLS}
Q. Wu, S. Zhang, H.-F. Song, M. Troyer, and A. A. Soluyanov, Comput. Phys. Commun. {\bf 224}, 405-416 (2017).

%Vanderbilt: screw symmetry protected (triple) Weyl points in Te
\bibitem{Vanderbilt_Te}
S. S. Tsirkin, I. Souza, and D. Vanderbilt, Phys. Rev. B {\bf 96}, 045102 (2017).

%Line node stuff
\bibitem{CHAN2016}
Y.-H. Chan, C.-K. Chiu, M. Y. Chou, and A. P. Schnyder, Phys. Rev. B {\bf 93}, 205132 (2016).
% Ca3P2 and other topological semimetals with line nodes and drumhead surface states
\bibitem{Vander1993}
D. Vanderbilt and R. D. King-Smith, Phys. Rev. B {\bf 48}, 4442 (1993).
%electric polarization as a bulk quantity and relation to the surface charge 

\bibitem{XU2017}
Q. Xu, R. Yu, Z. Fang, X. Dai, and H. Weng, Phys. Rev. B {\bf 95}, 045136 (2017).
%Topological nodal line semimetals in the CaP3 family of materials
\bibitem{QUAN2017}
Y. Quan, Z. P. Yin, and W. E. Pickett, Phys. Rev. Lett. {\bf 118}, 176402 (2017).
%Single nodal loop of accidental degeneracies in minimal symmetry: triclinic CaAs3
%
\bibitem{GM_Vander}
%Chiral degeneracies and Fermi-surface Chern numbers in bcc Fe
D. Gos\'albez-Mart\'inez, I. Souza, and D. Vanderbilt, Phys. Rev. B {\bf 92}, 085138 (2015).

%%Green's Function Method%%
\bibitem{MPLSancho}
M. P. L\'opez Sancho, J. M. L\'opez Sancho, and J. Rubio, J. Phys. F: Met. Phys. {\bf 15}, 851-858 (1985).
%%End GFM%%

%shifted dirac cones in TIs in-plane magnetic field
\bibitem{ZYUZ11}
A. A. Zyuzin, M. D. Hook, and A. A. Burkov, Phys. Rev. B {\bf 83}, 245428 (2011).
\bibitem{PERSH12}
S. S. Pershoguba and V. M. Yakovenko, Phys. Rev. B {\bf 86}, 165404 (2012).
%end shifted Dirac cones

\bibitem{XWang2006}
%Ab initio calculation of the anomalous Hall conductivity by Wannier interpolation
X. Wang, J. R. Yates, I. Souza, and D. Vanderbilt, Phys. Rev. B {\bf 74}, 195118 (2006).

%Haldane paper: attachment of surface Fermi arcs to Bulk fermi surface
\bibitem{HALDANE}
F. D. M. Haldane, arXiv:1401.0529v1 (2014).


%question fass
%\bibitem{KARG16}
%M. Kargarian, M. Randeria, and Y.-M. Lu, Proc. Natl. Acad. Sci. U.S.A. {\bf 113}, 8648 (2016).
%Are the surface Fermi arcs in Dirac semimetals topologically protected?.
%\bibitem{KARG17}
%M. Kargarian, Y.-M. Lu, M. Randeria, arXiv:1712.03982 (2017).
%Deformation and stability of surface states in Dirac semimetals
%\bibitem{LE18}
%C. Le, X. Wu, S. Qin, Y. Li, R. Thomale, F. Zhang, and J. Hu, arXiv:1801.05719v1 (2018).

%WSM + DSM review:
%\bibitem{Armitage2018}
%N. P. Armitage, E. J. Mele, and A. Vishwanath, Rev. Mod. Phys. {\bf 90}, 015001 (2018).

%%%% Wannierization %%%%%%%%%%%%%%%%%%%%%%%

\end{thebibliography}
\end{document}